\documentclass[journal]{IEEEtran}
\usepackage{amsmath}
\usepackage{amssymb}
\usepackage{bm}
\usepackage{enumitem}
\usepackage{multirow}

\usepackage{booktabs}
\usepackage{xspace}
\newcommand{\sigmac}{\bm{\Sigma}^{(\bm{c})}}
\newcommand{\sigmao}{\bm{\Sigma}^{(\bm{o})}}
\newcommand{\sigmaf}{\bm{\Sigma}^{(\bm{o}^{\mathrm{(F0)}})}}
\newcommand{\prior}{\sigma_p}

\newcommand{\static}{$\mathcal{L}^{\mathrm{(s)}}$}
\newcommand{\staticdynamic}{$\mathcal{L}$}
\newcommand{\dynamic}{$\mathcal{L}^{\mathrm{(d)}}$}

\newcommand{\formse}{$\texttt{RMSE}_\mathrm{nat}$\xspace}
\newcommand{\focorr}{$\texttt{CORR}_\mathrm{nat}$\xspace}
\newcommand{\notermse}{$\texttt{RMSE}_\mathrm{note}$\xspace}
\newcommand{\notecorr}{$\texttt{CORR}_\mathrm{note}$\xspace}

\usepackage{threeparttable}

\ifCLASSINFOpdf
  \usepackage[pdftex]{graphicx}
\else
  \usepackage[dvipdfmx]{graphicx}
\fi

\usepackage{cite}
\ifCLASSINFOpdf
\else
\fi
\ifCLASSOPTIONcompsoc
 \usepackage[caption=false,font=normalsize,labelfont=sf,textfont=sf]{subfig}
\else
 \usepackage[caption=false,font=footnotesize]{subfig}
\fi
\usepackage{url}
\begin{document}
%
% paper title
% Titles are generally capitalized except for words such as a, an, and, as,
% at, but, by, for, in, nor, of, on, or, the, to and up, which are usually
% not capitalized unless they are the first or last word of the title.
% Linebreaks \\ can be used within to get better formatting as desired.
% Do not put math or special symbols in the title.
\title{Sinsy: A Deep Neural Network-Based Singing Voice Synthesis System}
%
%
% author names and IEEE memberships
% note positions of commas and nonbreaking spaces ( ~ ) LaTeX will not break
% a structure at a ~ so this keeps an author's name from being broken across
% two lines.
% use \thanks{} to gain access to the first footnote area
% a separate \thanks must be used for each paragraph as LaTeX2e's \thanks
% was not built to handle multiple paragraphs
%

\author{Yukiya~Hono,
        Kei Hashimoto,~\IEEEmembership{Member,~IEEE,}
        Keiichiro Oura,
        Yoshihiko Nankaku,~\IEEEmembership{Member,~IEEE,}\\
        and~Keiichi Tokuda,~\IEEEmembership{Fellow,~IEEE}% <-this % stops a space
\thanks{The authors are with Nagoya Institute of Technology.}% <-this % stops a space
\thanks{Manuscript received xxxxx xx, 2021; revised xxxxx xx, 2021.}}

% note the % following the last \IEEEmembership and also \thanks -
% these prevent an unwanted space from occurring between the last author name
% and the end of the author line. i.e., if you had this:
%
% \author{....lastname \thanks{...} \thanks{...} }
%                     ^------------^------------^----Do not want these spaces!
%
% a space would be appended to the last name and could cause every name on that
% line to be shifted left slightly. This is one of those "LaTeX things". For
% instance, "\textbf{A} \textbf{B}" will typeset as "A B" not "AB". To get
% "AB" then you have to do: "\textbf{A}\textbf{B}"
% \thanks is no different in this regard, so shield the last } of each \thanks
% that ends a line with a % and do not let a space in before the next \thanks.
% Spaces after \IEEEmembership other than the last one are OK (and needed) as
% you are supposed to have spaces between the names. For what it is worth,
% this is a minor point as most people would not even notice if the said evil
% space somehow managed to creep in.

% The paper headers
\markboth{Journal of \LaTeX\ Class Files,~Vol.~0, No.~0, August~2021}%
{Shell \MakeLowercase{\textit{et al.}}: Bare Demo of IEEEtran.cls for IEEE Journals}
% The only time the second header will appear is for the odd numbered pages
% after the title page when using the twoside option.
%
% *** Note that you probably will NOT want to include the author's ***
% *** name in the headers of peer review papers.                   ***
% You can use \ifCLASSOPTIONpeerreview for conditional compilation here if
% you desire.

% If you want to put a publisher's ID mark on the page you can do it like
% this:
%\IEEEpubid{0000--0000/00\$00.00~\copyright~2015 IEEE}
% Remember, if you use this you must call \IEEEpubidadjcol in the second
% column for its text to clear the IEEEpubid mark.

% use for special paper notices
%\IEEEspecialpapernotice{(Invited Paper)}

% make the title area
\maketitle

% As a general rule, do not put math, special symbols or citations
% in the abstract or keywords.
\begin{abstract}
  This paper presents Sinsy, a deep neural network (DNN)-based singing voice synthesis (SVS) system.
  In recent years, DNNs have been utilized in statistical parametric SVS systems, and DNN-based SVS systems have demonstrated better performance than conventional hidden Markov model-based ones.
  SVS systems are required to synthesize a singing voice with pitch and timing that strictly follow a given musical score.
  Additionally, singing expressions that are not described on the musical score, such as vibrato and timing fluctuations, should be reproduced.
  The proposed system is composed of four modules: a time-lag model, a duration model, an acoustic model, and a vocoder, and singing voices can be synthesized taking these characteristics of singing voices into account.
  To better model a singing voice, the proposed system incorporates improved approaches to modeling pitch and vibrato and better training criteria into the acoustic model.
  In addition, we incorporated PeriodNet, a non-autoregressive neural vocoder with robustness for the pitch, into our systems to generate a high-fidelity singing voice waveform.
  Moreover, we propose automatic pitch correction techniques for DNN-based SVS to synthesize singing voices with correct pitch even if the training data has out-of-tune phrases.
  Experimental results show our system can synthesize a singing voice with better timing, more natural vibrato, and correct pitch, and it can achieve better mean opinion scores in subjective evaluation tests.
\end{abstract}

% Note that keywords are not normally used for peerreview papers.

% For peer review papers, you can put extra information on the cover
% page as needed:
% \ifCLASSOPTIONpeerreview
% \begin{center} \bfseries EDICS Category: 3-BBND \end{center}
% \fi
%
% For peerreview papers, this IEEEtran command inserts a page break and
% creates the second title. It will be ignored for other modes.
\IEEEpeerreviewmaketitle

\section{Introduction}
% The very first letter is a 2 line initial drop letter followed
% by the rest of the first word in caps.
%
% form to use if the first word consists of a single letter:
% \IEEEPARstart{A}{demo} file is ....
%
% form to use if you need the single drop letter followed by
% normal text (unknown if ever used by the IEEE):
% \IEEEPARstart{A}{}demo file is ....
%
% Some journals put the first two words in caps:
% \IEEEPARstart{T}{his demo} file is ....
%
% Here we have the typical use of a "T" for an initial drop letter
% and "HIS" in caps to complete the first word.
\IEEEPARstart{S}{inging} voice synthesis (SVS) is a technique of generating singing voices from musical scores.
A unit-selection method~\cite{kenmochi-2007-vocaloid,bonada-2016-expressive} can automatically synthesize a singing voice by concatenating short waveform units selected from a database.
While such systems can provide good sound quality and naturalness in certain settings, it is impossible to guarantee that the units will always be connected smoothly.
Moreover, since it also tends to have limited flexibility, large databases are generally required to synthesize singing voices.

Statistical parametric SVS systems such as hidden Markov model (HMM)-based SVS systems~\cite{oura-2010-recent} have been proposed to avoid the problems described above.
The singing voice waveform is synthesized from the acoustic parameters predicted by a trained HMM, thereby requiring less data to construct a system compared to unit-selection systems.
However, HMM-based systems suffer from over-smoothing that degrades the naturalness of synthesized singing voices.

In recent years, deep neural networks (DNNs) have significantly improved in various speech processing tasks such as speech recognition~\cite{hinton-2012-deep}, speech synthesis~\cite{zen-2013-statistical,qian-2014-training}, and voice conversion~\cite{desai-2010-spectral}.
DNN-based SVS systems~\cite{nishimura-2016-singing,hono-2018-recent} have also been proposed and demonstrated their superiority over HMM-based ones.
A feed-forward neural network (FFNN) is utilized as an acoustic model to represent the mapping function between the musical score feature and the acoustic feature.
Recently, recurrent neural networks (RNNs) with long short-term memory (LSTM), convolutional neural networks (CNNs), and deep autoregressive (AR) models have been incorporated into SVS systems to model the acoustic features more appropriately~\cite{kim-2018-korean,nakamura-2020-fast,blaauw-2017-neural-2,yi-2019-singing}.
Trajectory training~\cite{hashimoto-2016-trajectory} and adversarial training~\cite{goodfellow-2014-generative} have also been incorporated into SVS systems to improve training criteria and achieve higher singing voice quality~\cite{hono-2018-recent,hono-2019-singing}.

In SVS systems, singing voices must be synthesized accurately following the input musical score.
Methods such as pitch normalization~\cite{nishimura-2016-singing} and data augmentation~\cite{mase-2010-hmm,blaauw-2017-neural-2} have been proposed for DNN-based SVS systems to generate fundamental frequency (F0) following the note pitch in the input musical score.
Vibrato is the periodic fluctuation of the pitch and is another essential point of modeling a singing voice.
Some systems model vibrato-like fluctuations as a part of the F0~\cite{kim-2018-korean,blaauw-2017-neural-2,yi-2019-singing}.
Our previous work~\cite{hono-2018-recent} separates the vibrato from the F0 and models it as sinusoidal parameters, enabling reproduction and control of the vibrato.
The temporal structure of a singing voice is heavily constrained by note length in a musical score, but the start timing of musical notes and a singing voice do not always match.
A framework with a time-lag model and a duration model has been proposed to determine the phone durations under note length constraints considering these timing fluctuations~\cite{hono-2018-recent}.
These techniques are essential for synthesizing a human-like natural singing voice.

When building SVS systems, a pitch correction technique is sometimes necessary to avoid generating out-of-tune singing voices.
Since DNN-based SVS is a statistical approach that tries to reproduce training data, it tends to generate an out-of-tune pitch if the training data contains out-of-tune phrases.
The pitch accuracy significantly impacts the subjective quality of the singing voices; thus, a technique is needed for synthesizing singing voices with an appropriate pitch from arbitrary training data, including such out-of-tune phrases.

Recently, TTS research fields have utilized state-of-the-art systems with sequence-to-sequence (seq-to-seq) acoustic models and neural waveform generation models to achieve the same naturalness as human speech~\cite{shen-2018-natural}.
Seq-to-seq models with attention mechanisms directly map input text or phonetic sequences to the acoustic features without using an external duration model.
Although some seq-to-seq models for end-to-end SVS have also been proposed~\cite{lee-2019-adversarially,angelini-2020-singing,blaauw-2020-sequence,gu-2020-bytesing,lu-2020-xiaoicesing,chen-2020-hifisinger,shi-2020-sequence}, unlike TTS, a duration informed attention network is mainly used because of singing-specific backgrounds.
For instance, the lengths of singing voices are generally longer than those of speech, and the amount of training data is insufficient.
With the growth of deep learning techniques, statistical parametric SVS has been attracting attention for its various applications; however, these systems require high stability and controllability in terms of both acoustic parameters and alignments.
Furthermore, since the synthesized singing voice strictly needs to be synchronized with the given musical score, it is not enough to apply the TTS-like end-to-end frameworks to the SVS systems.
There is still a high demand for pipeline systems with an external time-lag model and a duration model from these perspectives.

This paper presents our DNN-based SVS system, ``Sinsy.''
Our proposed system of this paper is an extension of our previous work~\cite{hono-2018-recent}.
All the components for synthesizing a singing voice from the analyzed score features are based on neural networks and incorporate novel techniques to better model a singing voice.
Our system has a singing-specific design:
1) The combination strategy with the time-lag model and the duration model predicts phoneme boundaries under note length constraints statistically.
2) The acoustic model has improved pitch and vibrato modeling and a better training criterion for considering dynamic features.
3) The PeriodNet~\cite{hono-2021-periodnet}, a non-AR neural vocoder with more robustness of pitch, is adopted.
4) Automatic pitch correction techniques are incorporated into our SVS system to synthesize singing voices with the correct pitch.
With these techniques, our proposed system can synthesize a high-fidelity singing voice waveform.

In the rest of this paper, Section~\ref{sec:related} reviews the conventional SVS system.
Section~\ref{sec:nnsvs} describes the overview of our proposed SVS system.
Section~\ref{sec:specific} introduces the proposed techniques for modeling pitch and vibrato.
Section~\ref{sec:pitch} describes our proposed automatic pitch correction methods for DNN-based SVS.
Section~\ref{sec:exp} presents the experimental evaluations.
Finally, Section~\ref{sec:conclusion} concludes this paper.

\section{Related Work}
\label{sec:related}
The usage of neural networks in SVS systems is similar to that in TTS systems.
The simplest way to apply DNNs to TTS systems is to use an FFNN as a deep regression model to map a linguistic feature sequence obtained by text to an acoustic feature sequence extracted from speech~\cite{zen-2013-statistical}.
A DNN-based SVS system also uses the DNN as the acoustic model; however, unlike TTS, feature vectors extracted from the musical score are used as the input instead of the linguistic feature.
Architectures such as RNNs, CNNs, and AR structures are used as acoustic models
for both TTS and SVS systems~\cite{fan-2014-tts,wang-2016-gating,wang-2017-autoregressive,kim-2018-korean,nakamura-2020-fast,yi-2019-singing,blaauw-2017-neural-2}.

The pitch of the synthesized singing voice must accurately follow the note pitch of the musical score even if the note pitch to be synthesized is outside the range of the training data.
A pitch normalization technique has been proposed for F0 modeling in DNN-based SVS~\cite{nishimura-2016-singing}.
In this technique, the differences between the log F0 sequence extracted from waveforms and the note pitch are modeled.
Recent studies~\cite{lu-2020-xiaoicesing,chen-2020-hifisinger} introduced a residual connection between note pitch and generated log F0, which can be said to be the same approach to pitch normalization.
Some systems~\cite{mase-2010-hmm,blaauw-2017-neural-2} utilize a data augmentation technique by pitch-shifting the training data.
However, this technique requires more training time due to the increased amount of training data, and it is difficult to reproduce the voice characteristics and singing styles that change according to the tone.
A post-processing strategy has also been proposed~\cite{yi-2019-singing}.
For this strategy, F0 should be modified for each voiced segment, which may generate a discontinuous F0 contour at the edge of the voiced segment.

Another unique characteristic of singing voices is that F0 includes periodic fluctuations due to vibrato.
In our previous study~\cite{hono-2018-recent}, we separated the vibrato from the original F0 sequence in advance and modeled with sinusoidal parameters.
This enables us to control the vibrato intensity and speed in the synthesis stage.
Some systems do not use the decomposed approach, and the F0 sequence with the vibrato component is directly modeled using neural networks such as RNNs and AR models~\cite{kim-2018-korean,blaauw-2017-neural-2,yi-2019-singing}.
This approach can be expected to reproduce more complex vibrato shapes that are difficult to represent by sinusoidal parameters.
However, there is a problem that the vibrato cannot be controlled during synthesis.

\begin{figure*}[t]
  \centering
  \includegraphics[width=0.9\hsize]{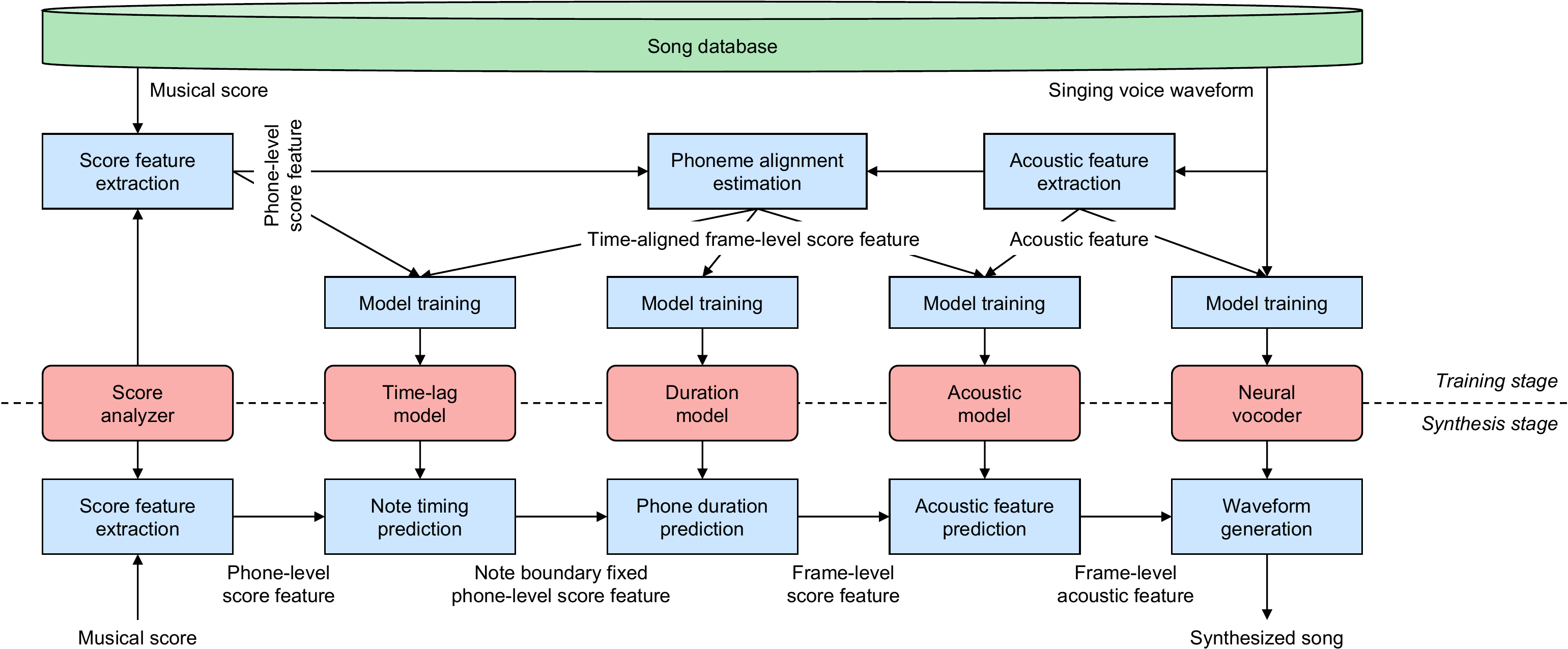}
  \vspace{-1mm}
  \caption{
    Overview of the DNN-based SVS system, ``Sinsy.''
    Our system consists of a score analyzer and four neural network-based models.
    Singing voice can be synthesized from a musical score via these modules.
    Proposed singing-specific techniques, pitch normalization, vibrato modeling, and automatic pitch correction are incorporated in the acoustic model.
  }
  \label{fig:svss}
\end{figure*}

Since DNN-based SVS systems are data-driven approaches, the quality of the synthesized singing voice depends on the quality and quantity of the training data.
Recent TTS systems often use over 20 hours of training data for a single-speaker model~\cite{shen-2018-natural} and even hundreds of hours for a multi-speaker model~\cite{zen-2019-libritts}.
However, unlike TTS, the amount of training data for SVS is often limited due to recording costs, annotation costs, and strict copyright issues in the music domain.
Thus, in most cases, 1-2 hours of a singing voice corpus are used.
Recently, there has been an attempt to utilize singing voice data mined from music websites as training data~\cite{ren-2020-deepsinger}.
However, the pitch of this mined data is not always correct, despite pitch accuracy significantly impacting the quality of a singing voice.
Even if the singing voice data is recorded for the training data, it may contain out-of-tune data due to various factors, such as a singer's skill, a song's tempo, and/or a melody's complexity.
Although F0 contours can be manually modified after being extracted from training data, it is difficult to correct while maintaining a human-like F0 trajectory and is impractical in terms of editing cost.
Therefore, there is demand for an automatic pitch correction technique in the SVS system.

Our system is based on our previous studies~\cite{nishimura-2016-singing,hono-2018-recent} and incorporates improved singing-specific techniques.
The skip connection of the note pitch improves acoustic feature estimation accuracy, particularly pitch.
The differences-based vibrato modeling can achieve more accurate reproduction of vibrato shape and can help control vibrato intensity.
Automatic pitch correction techniques are also introduced into the SVS system.
In addition, our system adopts the pitch robust neural vocoder PeriodNet.
As a result, our system can synthesize a more natural singing voice that follows a given score more accurately.

\section{DNN-Based SVS System}
\label{sec:nnsvs}
\subsection{Overview}

Figure~\ref{fig:svss} overviews our proposed DNN-based SVS system, ``Sinsy.''
This system consists of several models:
1) a time-lag model to predict the start timing of the notes,
2) a duration model to predict phoneme durations in each note,
3) an acoustic model to generate acoustic features based on the predicted phoneme timing, and
4) a vocoder to synthesize waveforms from generated acoustic features.
These models are based on the neural network.

Our system is composed of training and synthesis parts.
In the training part, score and acoustic features are extracted by score analysis and vocoder encoding, then each model is trained.
The score feature contains musical score information (e.g., lyrics, note keys, note lengths, tempo, dynamics, and slur), and the acoustic feature contains spectrum (e.g., mel-cepstral coefficients) and excitation parameters (e.g., F0).
Time-aligned score features are needed to train the time-lag model, the duration model, and the acoustic model.
Therefore, pre-trained hidden semi-Markov models (HSMMs) are used to estimate phoneme alignments~\cite{zen-2007-hidden}.

In the synthesis part, first, the score features are extracted from the musical score to be synthesized.
The time-lag model predicts each note's start timing, and the duration model predicts the phoneme duration in each note under the constraints of the note boundaries determined by the time-lag model.
The frame-level score feature sequence is obtained using these predicted boundaries and then fed into the acoustic model to predict the acoustic feature sequence.
Finally, the neural vocoder synthesizes a singing voice waveform.

Sinsy is a system for synthesizing singing voices from music scores.
We adopt MusicXML~\cite{Web-MusicXML-date} for representing musical scores that include lyrics.
The score analyzer extracts musical contexts from the input musical score and encodes them into the categorical and numerical features that are easy for neural networks to handle.
We use singing specific rich contexts, which are designed in our previous works~\cite{oura-2010-recent}.

\subsection{Acoustic Model}

In the SVS, the DNN-based acoustic model represents the mapping function from the score feature sequences to the acoustic feature sequences.
There is a correlation between the spectrum and the excitation parameters in the acoustic feature (e.g., mel-cepstral coefficients and F0).
Some studies have utilized a cascade structure to model this correlation~\cite{blaauw-2017-neural-2}.
In this work, a single neural network is used to model both spectrum and excitation parameters simultaneously, assuming that the correlation between them can be expressed inside the neural network.

The sequence of acoustic feature vectors $\bm{c}$ can be written in vector forms as follows:
\begin{align}
  \bm{c} &= [\bm{c}_1^\top, \ldots, \bm{c}_t^\top, \ldots, \bm{c}_T^\top]^\top\!,
\end{align}
where $\bm{c}_t$ is a $D$-dimensional static feature vector that can be represented by $\bm{c}_t = [c_t(1), c_t(2), \ldots, c_t(D)]^\top$, and $T$ is the number of frames included in a song.
The optimal static feature vector sequence $\hat{\bm{c}}$ is given by
\begin{align}
  \hat{\bm{c}} &= \arg\max_{\bm{c}} \mathcal{N} (\bm{c} \mid \bar{\bm{c}}, \sigmac), \label{eq:gen-static}
\end{align}
where $\mathcal{N}(\,\cdot\, |\, \bar{\bm{c}}, \sigmac)$ denotes the Gaussian distribution with a mean vector $\bar{\bm{c}}$ and a covariance matrix $\sigmac$.
In the SVS system, $\bar{\bm{c}}$ is obtained by feeding the score feature vector sequence into a trained neural network.
A covariance matrix $\sigmac$ is given by
\begin{align}
  \sigmac &= \mathrm{diag}[\sigmac_1, \ldots, \sigmac_t, \ldots, \sigmac_T].
\end{align}
In the DNN-based SVS, $\sigmac$ is usually independent of score features; thus, $\sigmac$ is a globally tied covariance matrix.
Training of the DNN aims to maximize the likelihood function $\mathcal{L}^{\mathrm{(s)}}$ given by
\begin{align}
  \mathcal{L}^{\mathrm{(s)}} &= \mathcal{N} (\bm{c} \mid \bar{\bm{c}}, \sigmac). \label{eq:loss-static}
\end{align}

Since a singing voice includes long tones, parameter discontinuity degrades the quality of the synthesized singing voice.
Our previous studies~\cite{nishimura-2016-singing,hono-2018-recent,hono-2019-singing} used the dynamic features to avoid this.
The acoustic feature sequence of static and their dynamic feature vectors\footnote{We use velocity and acceleration features as dynamic features.} $\bm{o}$ can be written in vector forms as follows:
\begin{align}
  \bm{o} &= [\bm{o}_1^\top, \ldots, \bm{o}_t^\top, \ldots, \bm{o}_T^\top]^\top\!,
\end{align}
where $\bm{o}_t$ consists of the static and the dynamic feature vectors $\bm{o}_t = [\bm{c}_t^\top, \Delta^{(1)}\bm{c}_t^\top, \Delta^{(2)}\bm{c}_t^\top]^\top$.
Relation between $\bm{o}$ and $\bm{c}$ can be represented by $\bm{o} = \bm{W}\bm{c}$, where $\bm{W}$ is a window matrix that extends $\bm{c}$ to $\bm{o}$.
An acoustic model is trained by maximizing the following objective function as
\begin{align}
  \mathcal{L}^{\mathrm{(d)}} &= \mathcal{N} (\bm{o} \mid \bar{\bm{o}}, \sigmao), \label{eq:loss-delta}
\end{align}
where $\bar{\bm{o}}$ and $\sigmao$ are a mean vector and a global tied covariance matrix that include the elements that correspond to dynamic features.
The optimal static feature vector sequence $\hat{\bm{c}}^{\mathrm{(d)}}$ is obtained from considering dynamic features by using the parameter generation algorithm~\cite{tokuda-2000-speech} as follows:
\begin{align}
  \hat{\bm{c}}^{\mathrm{(d)}} &= \arg\max_{\bm{c}} \mathcal{N} (\bm{o} \mid \bar{\bm{o}}, \sigmao) = \arg\max_{\bm{c}} \mathcal{N} (\bm{W}\bm{c} \mid \bar{\bm{o}}, \sigmao).
  \label{eq:gen-delta}
\end{align}

Although the parameter generation algorithm can generate a smooth acoustic feature sequence, the computational cost at the synthesis stage increases.
A recent study~\cite{nakamura-2020-fast} introduced a different approach that considers dynamic features only during training.
In this approach, the objective function that considers the dynamic features can be written as
\begin{align}
  \mathcal{L} = \mathcal{N} (\bm{o} \mid \bm{W}\bar{\bm{c}}, \sigmao). \label{eq:loss-stadelta}
\end{align}
Since the output of the DNN contains only static feature vector $\bar{\bm{c}}$, the optimal static feature sequence can be obtained by \eqref{eq:gen-static} without the parameter generation algorithm in the synthesis stage.
The average length of a singing voice is longer than that of speech, and the generation time is sometimes a problem.
Thus, we utilize this approach for our system to generate a smooth parameter sequence without increased computational costs in the synthesis stage.

\subsection{Time-Lag Model and Duration Model}
\label{sec:timing}

\begin{figure}[t]
  \centering
  \includegraphics[width=0.95\hsize]{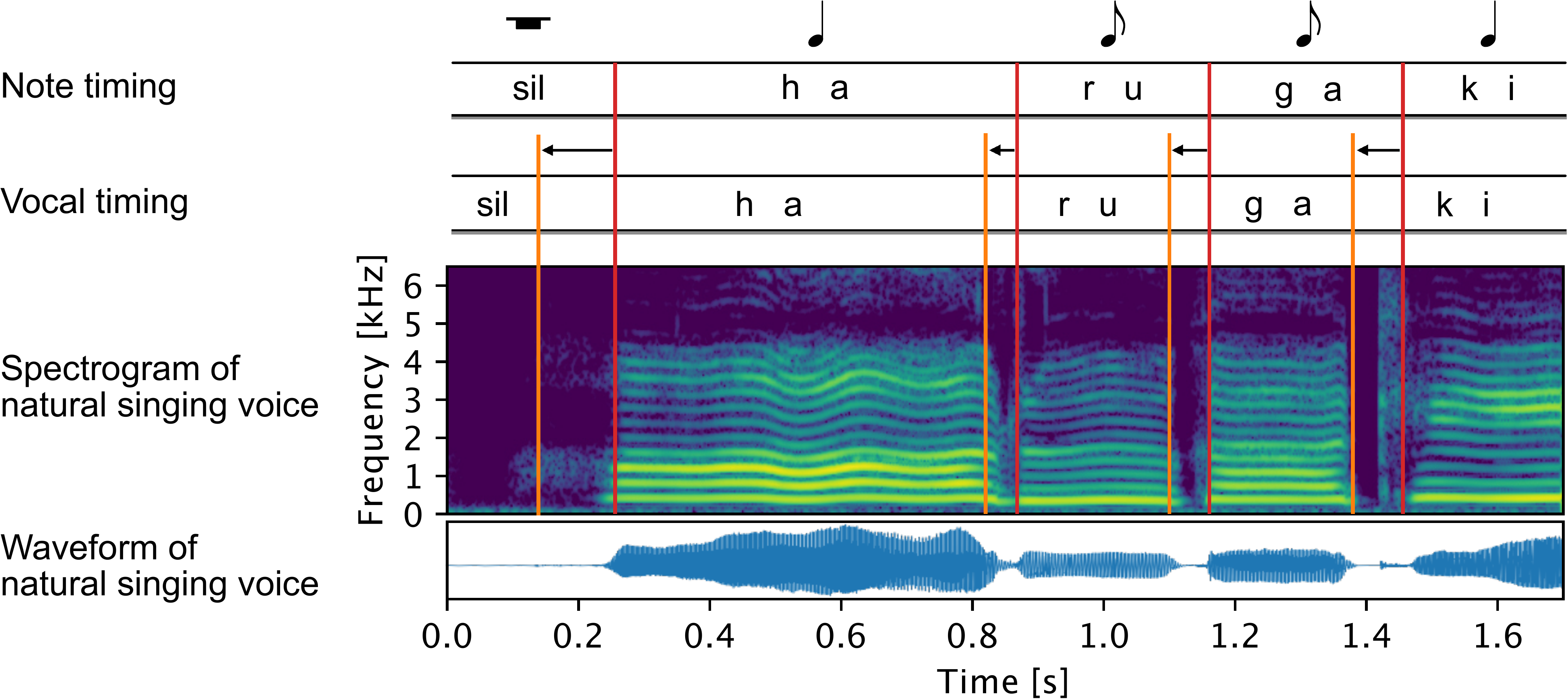}
  \vspace{-1mm}
  \caption{Example of time-lag.}
  \label{fig:time-lag}
\end{figure}

In SVS, the phoneme duration should be determined from the note length of the musical score since a singing voice is synthesized based on the tempo and rhythm of the music.
However, as Fig.~\ref{fig:time-lag} shows, humans generally tend to begin to utter consonants earlier than the absolute musical note onset timing.
In addition, note timing may be slightly advanced or delayed as part of an individual's singing technique, so the timing varies depending on the singer.
In this paper, we call the timing fluctuations caused by these factors ``time-lag'' and model them.

Two separated neural networks are used to model the time-lags and the phoneme durations.
We define a time-lag as the difference between the note timing of the musical score and the actual timing of a reference phoneme within a note.
To consider the time-lag in this work, we use the first vowel or silence for rests in each note as the reference phoneme instead of the first phoneme in the note.
This reference phoneme shifting is based on the tendency of humans to sing so that the vowel onset timing is closest to the note timing in the score, and preliminary experiments have confirmed its effectiveness.
Note that no additional annotation is needed in the training stage to train the time-lag and duration models because the actual phoneme timing of the singing voice can be obtained by a forced alignment using a well-trained HSMM~\cite{zen-2007-hidden}.

In the synthesis stage, first, the time-lag of each note is predicted using a trained time-lag model.
The sequences of note lengths obtained from the given musical score and predicted time-lags $\bm{L}$, $\hat{\bm{g}}$ can be written as follows:
\begin{align}
  \bm{L} &= [ L_1,\ldots,L_n,\ldots,L_N ], \\
  \hat{\bm{g}} &= [ \hat{g}_1,\ldots,\hat{g}_n,\ldots,\hat{g}_N ],
\end{align}
where $N$ is the number of notes included in a song.
Note that $\hat{g}_1$ is always zero since there is no need to shift the first note boundary.
Each adjusted note length $\hat{L}_n$ is obtained by
\begin{align}
  \hat{L}_n &= \left\{
  \begin{aligned}
    &L_n - \hat{g}_n + \hat{g}_{n+1}, & \quad(n<N) \\
    &L_n - \hat{g}_n. & \quad(n=N) \\
  \end{aligned}
  \right.
\end{align}

Next, the phoneme durations are predicted by a trained phoneme duration model and are normalized on a note-by-note basis so that the sums of the phoneme durations within each note match the adjusted note lengths.
The duration of the $k$-th phoneme in the $n$-th note is determined as follows:
\begin{align}
  \hat{d}_{nk} &= \hat{L}_n \cdot \mu_{nk} \biggl/\, \sum_{k=1}^{K_n} \mu_{nk} \biggl., \label{eq:dur-DNN}
\end{align}
where $K_n$ is the number of phonemes in the $n$-th note, and $\mu_{nk}$ is the output value of the DNN-based duration model at the $k$-th phoneme in the $n$-th note.

The phoneme duration of synthesized songs can be obtained by the above strategy, considering the time-lag by the neural network.
However, the distribution of phoneme durations differs greatly depending on the type of phonemes, such as consonants, vowels, and breaths.
For example, durations of vowels vary greatly depending on the note length, while those of consonants and breaths are barely affected by note length.
Note that the breath phoneme corresponds to the breath mark in the musical score, and its duration needs to be predicted because the actual duration cannot be obtained from the musical score.
Thus, it is not appropriate to fit the phoneme duration to the note length constraint by uniformly multiplying all the phonemes within a note by a constant as in \eqref{eq:dur-DNN}.

Therefore, we statistically estimate the adjusted phoneme duration by considering the variance of the phoneme duration.
This approach is based on constrained maximum likelihood estimation.
We use a mixture density network (MDN)~\cite{bishop-1994-mixture} to model the phoneme duration distribution.
Note that an MDN with one mixture component is used for simplification.
We assume that a single-mixture MDN has sufficient ability to represent the phoneme duration distribution.

The optimal phoneme duration sequence of $n$-th note $\hat{\bm{d}}_{n}^{\,\mathrm{(ML)}}$ is given by
\begin{align}
  \hat{\bm{d}}_n^{\,\mathrm{(ML)}} &= \arg\max_{\bm{\scriptstyle d}_n} \sum_{k=1}^{K_n} \log\mathcal{N} (d_{nk} \mid \mu_{nk}, \sigma^2_{nk}), \label{eq:dur-ML1}
\end{align}
subject to
\begin{align}
  \sum_{k=1}^{K_n} \hat{d}_{nk}^{\,\mathrm{(ML)}} &= \hat{L}_n, \label{eq:dur-ML2}
\end{align}
where $\mu_{nk}$, $\sigma_{nk}^2$ denote the mean and the variance of the $k$-th phoneme duration in the $n$-th note, and these are obtained from a trained MDN.

To obtain optimal phoneme durations under the constraint condition in \eqref{eq:dur-ML2}, we use the Lagrange multiplier method as follows:
\begin{align}
  F(d_{nk}, \rho_n) &= \sum_{k=1}^{K_n} \log \mathcal{N} (d_{nk} \mid \mu_{nk}, \sigma^2_{nk}) \nonumber \\
  &\qquad + \rho_n \left( \sum_{k=1}^{K_n} d_{nk} - \hat{L}_n \right), \label{eq:dur-lagrangeF}
\end{align}
where $\rho_n$ denotes the Lagrange multiplier.
Hence, the optimal duration of the $k$-th phoneme in the $n$-th note is obtained by
\begin{align}
  \hat{d}_{nk}^{\,\mathrm{(ML)}} &= \mu_{nk} + \rho_n \sigma^2_{nk}, \label{eq:dur-ML3}
\end{align}
where $\rho_n$ is given by
\begin{align}
  \rho_n = \left( \hat{L}_n - \sum_{k=1}^{K_n} \mu_{nk} \right) \biggl/\, \sum_{k=1}^{K_n} \sigma_{nk}^2 \biggl.. \label{eq:dur-ML4}
\end{align}

Our system assumes that each note is assigned one or more phonemes.
The note with the long sound symbol \mbox{``---''} is assigned the appropriate phoneme based on the previous note's lyric, using language-specific heuristic rules at the score analyzing process.
For example, in Japanese, the phoneme of a note with a long vowel symbol is obtained by duplicating a vowel in a previous note.
In English, the syllable nucleus allocated to the previous note is duplicated, and the phoneme after the duplicated syllable nucleus is shifted to the current note.
Since consecutive diphthongs due to duplication may degrade the continuity of a singing voice, we defined the duplication rules for diphthongs in our previous work~\cite{nakamura-2014-hmm}.

\begin{figure}
  \centering
  \includegraphics[width=0.9\hsize]{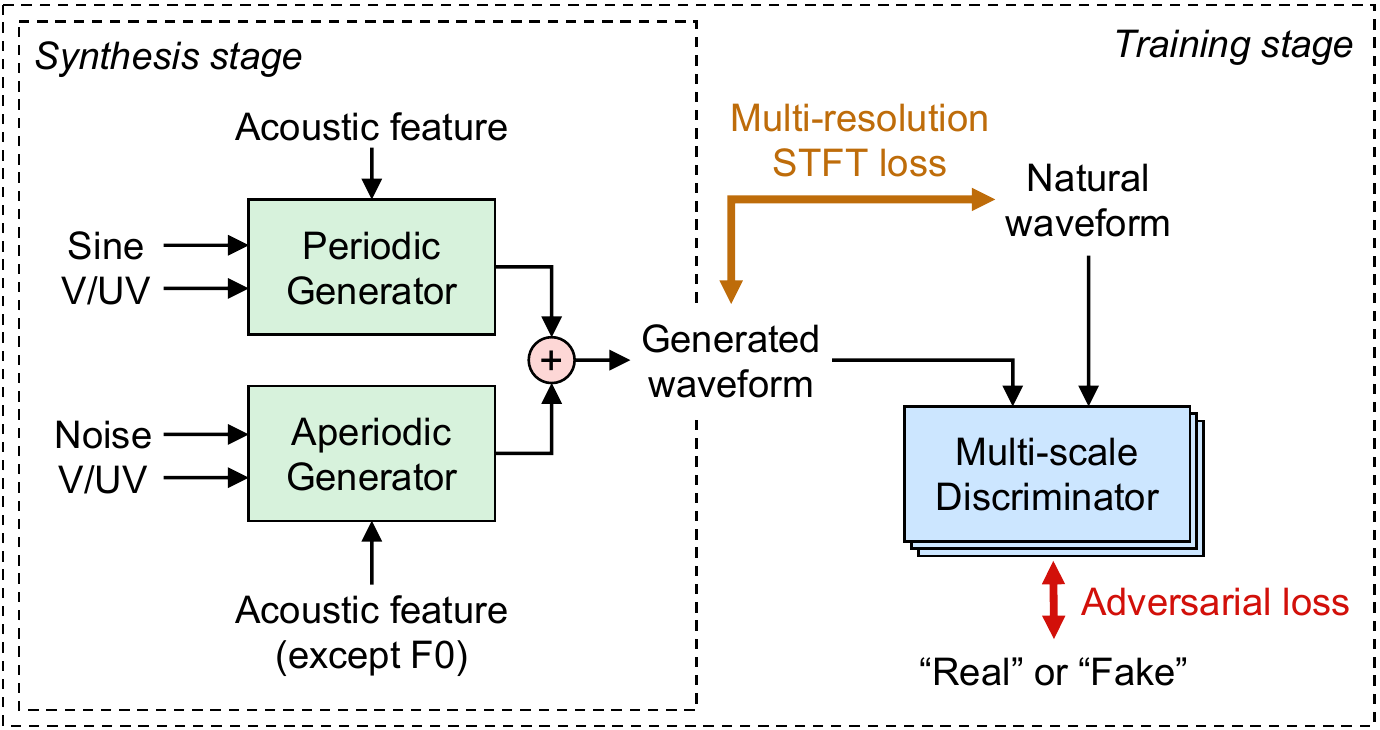}
  \vspace{-1mm}
  \caption{The overview of PeriodNet parallel model.}
  \label{fig:periodnet}
\end{figure}

\subsection{Neural Vocoder}

We use the PeriodNet~\cite{hono-2021-periodnet} as a vocoder to generate singing voice waveforms from acoustic feature sequences.
PeriodNet is a non-autoregressive neural vocoder with a structure that separates periodic and aperiodic components.
PeriodNet consists of two sub-generators connected in parallel or series and models a speech waveform based on the following two assumptions.
The first one is that the speech waveform can be represented as the sum of the periodic and aperiodic waveform.
The second one is that periodic and aperiodic waveforms of speech can be easily generated from an explicit periodic signal with autocorrelation (such as sinusoidal signal) and an explicit aperiodic signal without one (such as noise), respectively.
The parallel or series structure helps to model speech waveform with periodic and aperiodic components more appropriately and improves the robustness of the input acoustic features, especially F0.
SVS systems require the ability to generate high-quality singing voice waveforms even if the input pitch is outside the range of training data.
PeriodNet is highly suitable for the neural vocoder in SVS systems because it has superior reproducibility of accurate pitch and breath sounds.

This work adopts the parallel model (PM2 in~\cite{hono-2021-periodnet}), as shown in Fig.~\ref{fig:periodnet}.
A periodic generator takes an explicit periodic signal and an aperiodic generator that takes an aperiodic signal.
A sample-level voiced/unvoiced (V/UV) signal is also fed into both generators, and the periodic signal can be generated from F0 predicted by the acoustic model in the synthesis stage.
Both periodic and aperiodic generators adopt WaveNet-like architecture, which has a stack of non-causal convolution layers, and are conditioned on the acoustic feature.
To obtain the robustness of pitch, we exclude the F0 sequence from the condition of the aperiodic generator.
PeriodNet is trained by an adversarial training framework using a multi-scale discriminator along with a multi-resolution short-time Fourier transform (STFT) auxiliary loss.

\section{Accurate and Expressive Pitch Modeling for the DNN-Based SVS System}
\label{sec:specific}

This section describes the singing-specific techniques of accurately modeling pitch, including vibrato for the DNN-based SVS.

\subsection{Pitch Normalization}
\label{sec:pitch-norm}

The corpus-based nature of statistical parametric SVS approaches makes their performance highly dependent on the training data.
It is challenging to express contextual factors that rarely appear in training data.
Hence, DNN-based SVS systems should be trained using a database that contains various contextual factors to synthesize high-quality singing voices.
In particular, since the prediction accuracy of F0 significantly affects the quality of the synthesized singing voice, the pitch must be covered correctly.
However, it is almost impossible to cover all possible contextual factors because a singing voice is affected by a large number of factors, such as lyrics, key, dynamics, note duration, and note pitch.

A musical note-level pitch normalization technique for DNN-based SVS systems was proposed in our previous work~\cite{nishimura-2016-singing} to address the aforementioned problem.
In that technique, the F0 sequence extracted from the natural waveforms is not modeled directly but as a difference from the note pitch determined by musical notes in the score.
Therefore, the acoustic model only needs to predict the human bias from the note pitch.
This technique enables DNN-based SVS systems to synthesize singing voices that contain arbitrary pitches, including unseen ones.
However, modeling the difference of F0 remains a challenge because F0 in unvoiced frames and the note pitch in the musical rest are unmeasurable.
Thus, all unvoiced frames and musical rests in the musical score are linearly interpolated and modeled as voiced frames.

\begin{figure}
  \centering
  \includegraphics[width=0.8\hsize]{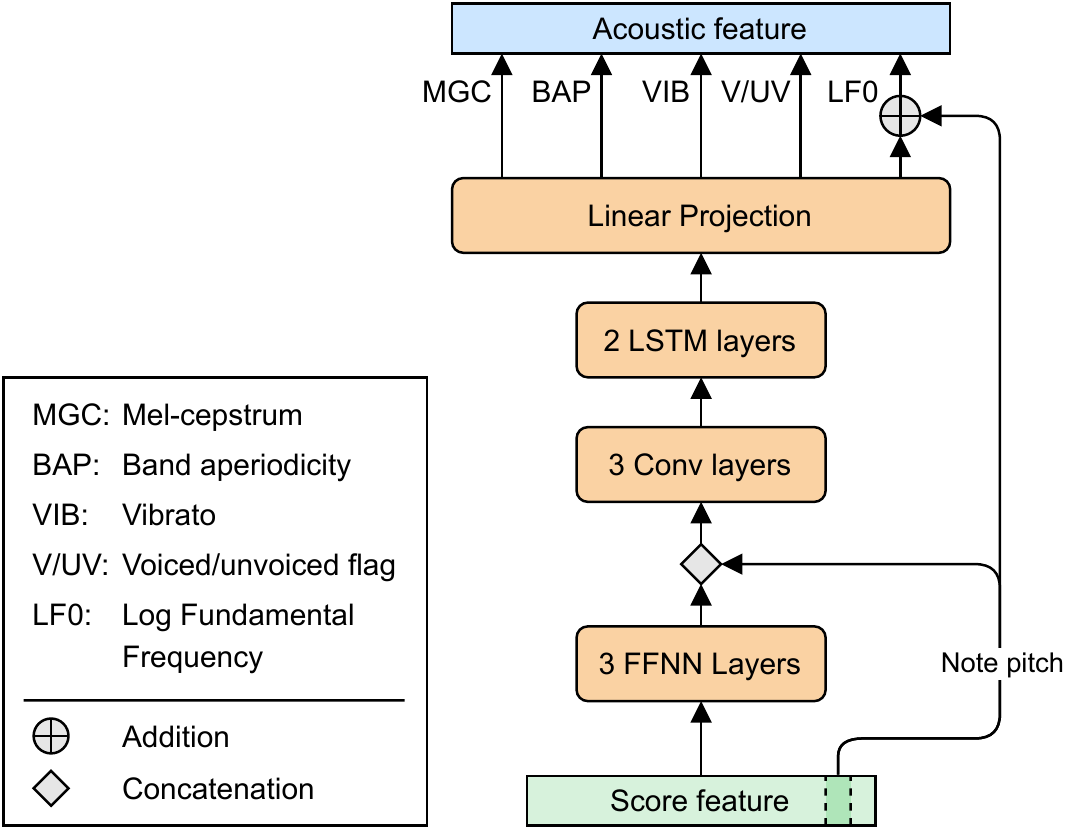}
  \caption{Architecture of acoustic model in our system.}
  \label{fig:acoustic_model}
\end{figure}

Figure~\ref{fig:acoustic_model} shows the architecture of the acoustic model with the pitch normalization technique.
In the pitch normalization technique, a predicted log F0 sequence $\bar{\bm{c}}^{(\mathrm{F0})}$ is represented by using the note pitch sequence $\bm{p} = [p_1, p_2, \ldots, p_T]^\top$ and the output mean parameter sequence $\bm{\mu} = [\mu_1, \mu_2, \ldots, \mu_T]^\top$ as follows:
\begin{align}
  \bar{\bm{c}}^{(\mathrm{F0})} = \bm{p} + \bm{\mu}. \label{eq:pitch-norm}
\end{align}
Note that we use log F0, the log scale of F0.
The note pitch sequence $\bm{p}$ to be added can be obtained from the input score features sequence.

A note pitch transition greatly influences the F0 trajectory.
Therefore, we add a skip connection between the input note pitch and a hidden layer of the acoustic model to deliver the note pitch inside the acoustic model, motivated by~\cite{nakamura-2020-fast}.
This helps transmit the note pitch information efficiently and predict the residual component between log F0 and the note pitch.

\subsection{Vibrato Model}

Generating an expressive F0 contour for a singing voice is also challenging.
Vibrato is one of the important singing techniques, and the timing and intensity of vibrato vary from singer to singer.
Thus, it should be modeled even though it cannot be explicitly described in the musical score.
Some acoustic models with a recurrent or AR structure can model F0 with vibrato directly~\cite{blaauw-2017-neural-2,kim-2018-korean,yi-2019-singing}.
However, it does not enable explicit control of vibrato.
Here, we assume that vibrato is a periodic fluctuation of the F0 contour and introduce two explicit vibrato modeling methods.

\subsubsection{Sine-Based Vibrato Modeling}
\label{sec:sine-vib}

One method for modeling vibrato is to express periodic fluctuations with sinusoidal parameters~\cite{hono-2018-recent}.
The vibrato $v(\cdot)$ of the $t$ frame in the $i$-th vibrato section $[t_i^{(s)},t_i^{(e)}]$ can be defined as
\begin{align}
  v\bigl(m_a(t), m_f(t), i\bigr) &= m_a(t) \sin\Bigl(2\pi m_f(t) f_s \bigl(t-t_i^{(s)}\bigr)\!\Bigr),
\end{align}
where $m_a(t)$ is the amplitude of vibrato in cents, $m_f(t)$ is the frequency of vibrato in Hz, and $f_s$ is the frame shift in seconds.
These parameters can be obtained from the original F0 sequence with an estimation algorithm~\cite{nakano-2006-automatic}.
In this work, the vibrato amplitude and frequency parameters are extracted based on the intersection points between the original F0 and a median-smoothed F0.
Two-dimensional parameters, $m_a(t)$ and $m_f(t)$, are added to the acoustic feature vector and modeled by a DNN along with the spectral and excitation parameters.
The vibrato parameters are unobserved outside the vibrato sections.
Thus, these parameters are interpolated in the same manner as the F0 sequence and are modeled along with an additional binary flag to determine the vibrato/non-vibrato frames.

\subsubsection{Difference-Based Vibrato Modeling}
\label{sec:diff-vib}

In this paper, we propose another method of modeling the vibrato component separated from the F0.
The vibrato component is defined as the difference between the original F0 sequence and the smoothed F0 sequence.
This difference-based vibrato component is a one-dimensional continuous feature and is directly modeled by the acoustic model that can model time series data such as RNN without an extra binary flag representing vibrato/non-vibrato frames.
This method has the advantage of generating more complex vibrato shapes given no assumption of the vibrato shape.
In particular, smoother vibrato can be obtained because the start and end of the vibrato are determined by the value of the difference rather than the binary flag.
Furthermore, it is possible to control the vibrato intensity by changing the difference values, unlike the method of not separating the vibrato component explicitly.

\section{Automatic Pitch Correction}
\label{sec:pitch}

The singing voice becomes out of tune when the pitch of a singing voice deviates from that of the musical score.
Therefore, we introduce two automatic correction techniques to prevent synthesized singing voices from becoming out of tune: prior distribution of pitch and pseudo-note pitch.

\subsection{Prior Distribution of Pitch}

In pitch normalization, the difference between log F0 and note pitch is modeled by \eqref{eq:pitch-norm}.
Here, we correct the out-of-tune phrases by giving a prior distribution to the pitch normalization training, assuming that the difference follows a zero-mean Gaussian distribution.

The prior distribution of pitch is given as
\begin{align}
  P(\bm{\mu}) = \mathcal{N}(\bm{\mu} \mid \bm{v}, \bm{S}),
\end{align}
where $\bm{v}$ and $\bm{S}$ correspond to the mean and the variance of prior Gaussian distribution.
Note that $\bm{\mu}$ does not contain a vibrato component because the pitch should be corrected while maintaining vibrato.
Here $\bm{v} = \bm{0}$ since the prior distribution we assume always represents the difference between the F0 extracted from the natural waveforms and note pitch in the musical score.
The variance $\bm{S}$ works as a parameter that controls the intensity of pitch correction.
The smaller element of $\bm{S}$ is, the stronger the out-of-tune pitch is corrected.
In this work, we assume that the variance always takes the fixed value $\prior^2$.
In the training part, an objective function in terms of F0 is defined as
\begin{align}
  \mathcal{L}^{\mathrm{(F0)}} &= \mathcal{N}(\bm{o}^{\mathrm{(F0)}} \mid \bm{W}(\bm{p} + \bm{\mu}), \sigmaf) \, \mathcal{N}(\bm{\mu} \mid \bm{v}, \bm{S}), \label{eq:loss-pitch}
\end{align}
where $\bm{o}^{\mathrm{(F0)}}$ and $\sigmaf$ are a sequence of feature vectors and a covariance matrix in terms of log F0 and their dynamic features.

\begin{figure}[t]
  \centering
  \includegraphics[width=0.95\hsize]{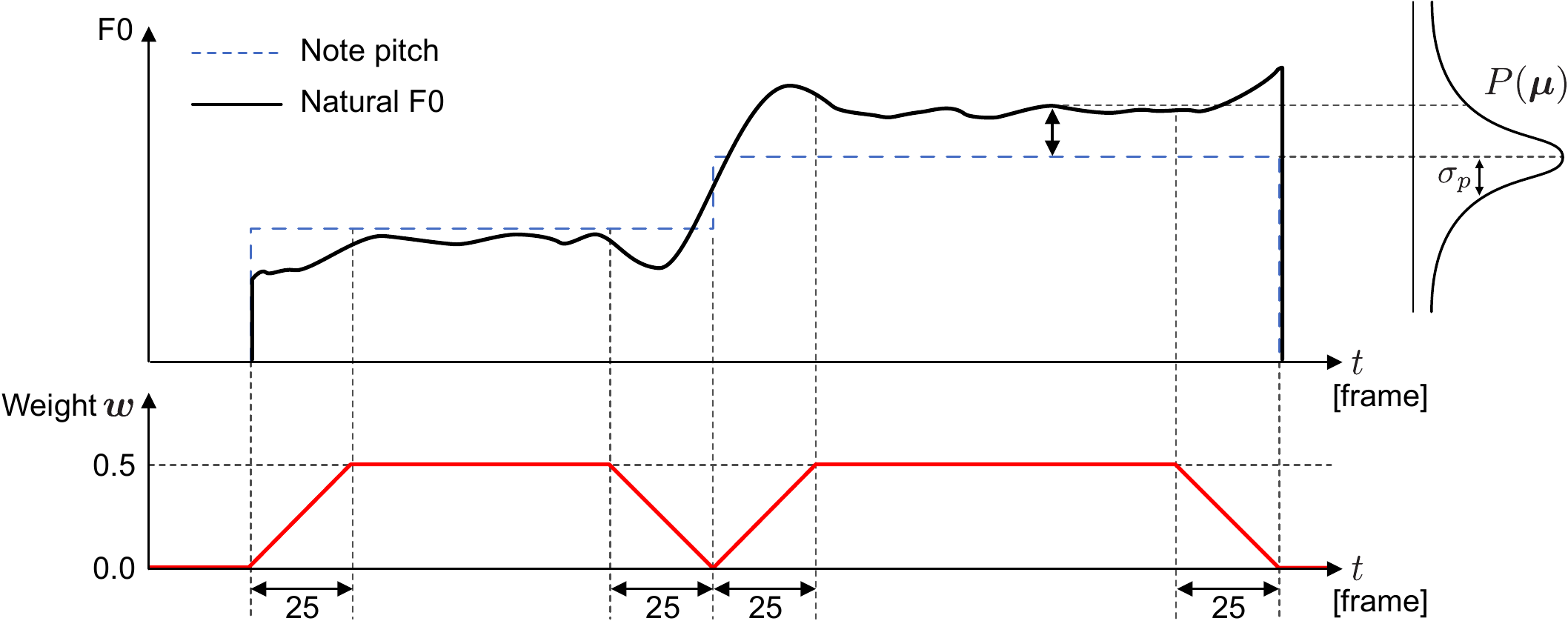}
  \vspace{-1mm}
  \caption{Automatic pitch correction by prior distribution with dynamic weight vector.}
  \label{fig:dweight}
\end{figure}

In unique phenomena of a singing voice such as overshooting\footnote{Overshooting is a pitch deflection exceeding the target note after a note change.} and preparation\footnote{Preparation is a pitch deflection in the direction opposite to a note change that can be seen just before the note change.}, F0 contours are deflected from the target notes before or after a note changes.
Thus, it is not always optimal to correct the pitch with the same intensity at all frames.
Furthermore, it sounds rather unnatural if the difference between the F0 and the note pitch becomes too small.
Therefore, we introduce a dynamic weight vector $\bm{w} = [w_1, w_2, \ldots, w_T]$ whose values are changed based on the note position into~\eqref{eq:loss-pitch}:
\begin{align}
  \mathcal{L}^{\mathrm{(F0)}} &= \mathcal{N}(\bm{o}^{\mathrm{(F0)}} \mid \bm{W}(\bm{p} + \bm{\mu}), \sigmaf) \, \mathcal{N}(\bm{\mu} \mid \bm{v}, \bm{S})^{\bm{w}} \nonumber \\
  &= \mathcal{N}(\bm{o}^{\mathrm{(F0)}} \mid \bm{W}(\bm{p} + \bm{\mu}), \sigmaf) \prod_{t=1}^T \mathcal{N}(\mu_t \mid 0, \prior^2)^{w_t}, \label{eq:loss-dpitch}
\end{align}
where $T$ denotes the number of frames.
The values of the weight vector increase or decrease at the beginning or end of notes, as shown in Fig.~\ref{fig:dweight}.
In this paper, the maximum of $\bm{w}$ takes $0.5$, and the width of increasing and decreasing is set to 25 frames.

\subsection{Pseudo-Note Pitch}

One cause of the phrases being out of tune is that there is a difference between an assumed note pitch by the singer and the correct note pitch.
By training the acoustic model using the pseudo-note pitch that takes these differences into account, the singing voice should be synthesized with the correct pitch by using the original note pitch during synthesis.
We propose two different approaches to obtaining the pseudo-note pitch.

\subsubsection{Heuristic Pseudo-Note Pitch}
\label{sec:pitch-heur}

The pseudo-note pitch is heuristically calculated from the flat part of the F0 sequence to express the singer-assumed note pitch during singing.
In the training stage, the pitch normalization technique is applied using this pseudo-note pitch.
In the synthesis stage, it is possible to synthesize the singing voice with the correct pitch by using the original note pitch instead of the pseudo-note pitch.
An example of the heuristic pseudo-note pitch will be shown in Section~\ref{sec:exp-pitch}.

\subsubsection{Pitch Bias-Based Pseudo-Note Pitch}
\label{sec:pitch-bias}

We introduce additional trainable parameters, note-level pitch bias, to represent the difference between the singer-assumed note pitch and the correct note pitch.
In this approach, the pseudo-note pitch is defined as the sum of the original note pitch given by musical score and the note-level pitch bias.
The pitch bias should absorb the average pitch shift in each note seen in the out-of-tune phrases.
In the training stage, the predicted log F0 sequence is defined as
\begin{align}
  \bar{\bm{c}}^{(\mathrm{F0})} = \bm{p} + \bm{\mu} + \bm{b}, \label{eq:pitch-bias}
\end{align}
where $\bm{b} = [b_1, b_2, \ldots, b_T]^\top$ is a sequence of the trainable pitch bias.
The bias parameters are assigned to each note (except for musical rests), and $\bm{b}$ is obtained by duplicating each parameter according to a corresponding note length.
Note that the bias values in musical rests are obtained by linearly interpolating them in the same manner as the original note pitch in pitch normalization described in Section~\ref{sec:pitch-norm}.
Hence, the total number of trainable bias parameters is equal to the total number of musical notes in the training data.
These bias parameters can be trained using a back-propagation in the same fashion as the other model parameters without any additional loss function.
In the synthesis stage, the F0 with the correct pitch can be generated by fixing the bias value to zero.

\section{Experiments}
\label{sec:exp}

This section evaluates the effectiveness of the proposed system in terms of the combination strategy with the time-lag model and the duration model, improved acoustic feature modeling, and automatic pitch correction techniques.

\subsection{Experimental Conditions}
\label{sec:exp-cond}

In this experiment, 70 Japanese children's songs (total: 70 min) by a female singer were used.
Sixty songs were used for training, and the rest were used for testing.
Singing voice signals were sampled at 48 kHz and windowed with a 5-ms shift.
The acoustic feature consisted of 0-th through 49-th mel-cepstral coefficients, log F0 value, 0-th through 24-th mel-cepstral analysis aperiodicity measures, and vibrato parameters.
In order to reduce an error of F0 extraction, voting results from three F0 estimators were used as F0 of acoustic features~\cite{sawada-2016-nitech}.
Mel-cepstral coefficients were extracted from the smoothed spectrum analyzed by WORLD~\cite{morise-2016-world}.
Two types of explicit vibrato parameters were used: 2-dimensional sinusoidal-based parameters that consisted of amplitude and frequency parameters and a 1-dimensional vibrato component that represented the difference between the original log F0 and the smoothed log F0.
The score feature was obtained by analyzing the musical score, and the contextual factor we used followed our previous work~\cite{oura-2010-recent}.

Five-state, left-to-right, no-skip HSMMs were used to obtain the time alignment of the score features and the acoustic features for training the DNN-based time-lag, duration, and acoustic models.
The decision tree-based context clustering technique was separately applied to distributions for the spectrum, excitation, aperiodicity, and state duration.
The spectrum and aperiodicity stream were modeled with single multivariate Gaussian distributions.
The excitation stream was modeled with multi-space probability distribution HSMMs (MSD-HSMMs)~\cite{tokuda-1999-hidden} that each consisted of a Gaussian distribution for ``voiced'' frames and a discrete distribution for ``unvoiced'' frames.
The duration stream was modeled with single Gaussian distributions.
The minimum description length (MDL) criterion was employed to control the size of the decision trees for context clustering~\cite{shinoda-1997-acoustic}.

\subsection{Objective Evaluation of Time-Lag Modeling and Duration Modeling}
\label{sec:exp-timing}

An objective evaluation experiment was conducted to compare the prediction accuracy of note timing and phoneme duration.
In this experiment, the following three methods were compared.
\begin{itemize}
  \setlength{\parskip}{1mm}
  \setlength{\itemsep}{1mm}
  \item \textbf{DT}:\;Conventional method of predicting phoneme boundaries using the decision tree-based clustered context-dependent time-lag model and duration model in an HMM-based SVS system~\cite{saino-2006-hmm}.
  \item \textbf{DNN}:\;Time-lag and phoneme duration were modeled by DNNs, and final phoneme boundaries were determined according to \eqref{eq:dur-DNN}.
  \item \textbf{DNN+ML}:\;Time-lag and phoneme duration were modeled by DNN and single MDN, and final phoneme boundaries were determined using constrained maximum likelihood estimation with \eqref{eq:dur-ML3}.
\end{itemize}

In \textbf{DT}, the sizes of the decision trees were determined by the MDL criterion.
In \textbf{DNN} and \textbf{DNN+ML}, the input feature of DNN was an 824-dimensional feature vector consisting of 734 binary features for categorical linguistic contexts (e.g., the current phoneme identity) and 90 numerical features for numerical contexts (e.g., the number of phonemes in the current syllable).
In \textbf{DNN}, the outputs of the time-lag model and duration model are one-dimensional numerical values.
In \textbf{DNN+ML}, the time-lag model outputs one-dimensional numerical values, and the duration model outputs the mean and variance of the one-dimensional phoneme duration distribution.
In \textbf{DNN} and \textbf{DNN+ML}, the architecture of the time-lag model was three hidden layers with 32 units per layer, and that of the duration model was three hidden layers with 256 units per layer.
The sigmoid activation function was used in the hidden layers, and the linear activation function was used in the output layer.
The weights of the DNNs and the MDN were initialized randomly, then the DNNs were optimized by minimizing the mean squared error, and the MDN was optimized by maximizing the likelihood.
In the training phase, the Adam optimizer~\cite{kingma-2014-adam} was adopted for all neural networks.

The root mean square errors (RMSEs) of the note and the phoneme duration and Pearson correlations (CORRs) of the note and the phoneme duration were used as the objective evaluation metrics.
Note that the phoneme boundaries of forced alignment obtained by trained HMMs were used as the correct phoneme boundaries.

\begin{table}[!t]
  \caption{Object Evaluation of Timing Prediction Accuracy}
  \label{tbl:obj-timing}
  \centering
  \begin{tabular}{l|ccc}
    \toprule
    Method & \textbf{DT} & \textbf{DNN} & \textbf{DNN+ML} \\
    \midrule\midrule
    Note duration-RMSE [frame]    & 15.78 & \textbf{12.75} & \textbf{12.75} \\
    Phoneme duration-RMSE [frame] & 13.32 & 11.23 & \textbf{10.94} \\
    \midrule
    Note duration-CORR            & 0.9742 & \textbf{0.9780} & \textbf{0.9780} \\
    Phoneme duration-CORR         & 0.9719 & 0.9757 & \textbf{0.9767} \\
    \bottomrule
  \end{tabular}
\end{table}

\begin{table*}
  \centering
  \begin{threeparttable}
    \caption{Results of Objective Evaluation of Acoustic Models}
    \label{tbl:obj-acoustic}
    \centering
    \begin{tabular}{c|cccc|ccccc}
      \toprule
      System & \multicolumn{4}{c|}{System Details} & MCD & F0-RMSE & F0+Vib-RMSE & \multirow{2}{*}{F0-CORR} & \multirow{2}{*}{F0+Vib-CORR} \\
      Index & Pitch norm.\tnote{\!a} & Skip connect.\tnote{\!b} & Vibrato\tnote{~\!c} & Criterion\tnote{~\!d}\hspace{1mm} & [dB] & [cent] & [cent] & & \\
      \midrule\midrule
      \textbf{System~1} & $\checkmark$ & $\checkmark$ & Diff-based & \staticdynamic & \textbf{5.423} & 74.00 & 80.96 & \textbf{0.9713} & \textbf{0.9647} \\
      \midrule
      \textbf{System~2} & & & Diff-based & \staticdynamic & 5.644 & 264.07 & 265.71 & 0.6689 & 0.6677 \\
      \textbf{System~3} & $\checkmark$ & & Diff-based & \staticdynamic & 5.627 & 79.80 & 86.69 & 0.9653 & 0.9585 \\
      \midrule
      \textbf{System~4} & $\checkmark$ & $\checkmark$ & Sine-based & \staticdynamic & 5.456 & \textbf{72.91} & 85.12 & 0.9712 & 0.9592 \\
      \textbf{System~5} & $\checkmark$ & $\checkmark$ & N/A & \staticdynamic & 5.439 & - & \textbf{80.95} & - & 0.9635 \\
      \midrule
      \textbf{System~6} & $\checkmark$ & $\checkmark$ & Diff-based & \static & 5.462 & 73.49 & 82.00 & 0.9712 & 0.9636 \\
      \textbf{System~7} & $\checkmark$ & $\checkmark$ & Diff-based & \dynamic & 5.445 & 74.16 & 82.48 & 0.9709 & 0.9633 \\
      \bottomrule
    \end{tabular}
    \begin{tablenotes}[para,flushleft,online,normal]
      \item[a]
      {\footnotesize Pitch normalization described in Section~\ref{sec:pitch-norm},}
      \item[b]
      {\footnotesize Skip connection described in Section~\ref{sec:pitch-norm},}
      \item[c]
      {\footnotesize ``Sine-based'' denotes sine-based vibrato modeling described in Section~\ref{sec:sine-vib}, and ``Diff-based'' denotes the difference-based vibrato modeling described in Section~\ref{sec:diff-vib}.}
      \item[d]
      {\footnotesize Trainig criteria \staticdynamic, \static, and \dynamic are given by \eqref{eq:loss-stadelta}, \eqref{eq:loss-static}, and \eqref{eq:loss-delta}, respectively.}
    \end{tablenotes}
  \end{threeparttable}
\end{table*}

The experimental results are listed in Table~\ref{tbl:obj-timing}.
It can be seen that both \textbf{DNN} and \textbf{DNN+ML} performed better in predicting note and phoneme boundaries than \textbf{DT}.
This result indicates the effectiveness of replacing the decision tree-based clustered models with DNN-based models.
Also, comparing \textbf{DNN+ML} with \textbf{DNN} in terms of phoneme duration prediction accuracy, \textbf{DNN+ML} outperformed \textbf{DNN}.
This suggests that constrained maximum likelihood estimation of note lengths with consideration of variances helps fit the phoneme durations.

\subsection{Comparison of Acoustic Feature Modeling}
\label{sec:exp-acoustic}

Objective and subjective evaluations were conducted to compare the acoustic models in terms of pitch normalization, skip connection of the note pitch, vibrato modeling, and the training criterion.
We used the seven systems shown in Table~\ref{tbl:obj-acoustic}.

The input feature vector for the acoustic models was an 844-dimensional feature vector with the 824-dimensional feature vector in Section~\ref{sec:exp-timing} and a 20-dimensional additional feature vector that included duration features.
The output feature vector for the acoustic models consists of mel-cepstral coefficients, log F0 value, mel-cepstral analysis aperiodicity measures, vibrato parameters (except \textbf{System~5}), voiced/unvoiced binary value, and vibrato/non-vibrato binary value (only \textbf{System~4}).
In \textbf{System~7}, the dynamic features (velocity and acceleration features) were also included in the output feature vector.
A single network that modeled all acoustic features simultaneously was trained by using the Adam optimizer~\cite{kingma-2014-adam}.
The architecture of the acoustic models was the stack of three fully connected layers with 2048 hidden ReLU units, three convolution blocks each containing a 1D convolutional layer with 1024 filters, batch normalization~\cite{ioffe-2015-batch} and ReLU activations, two bidirectional LSTMs containing 512 units (256 in each direction), and a linear projection layer.
The same pre-trained PeriodNet~\cite{hono-2021-periodnet} neural vocoder was used to reconstruct singing waveforms from generated acoustic features in all systems.

Mel-cepstral distortion (MCD) [dB], RMSEs of log F0 without/with vibrato (F0-RMSE and F0+Vib-RMSE) [cent], CORRs of log F0 without/with vibrato (F0-CORR and F0+Vib-CORR) were used to objectively evaluate the performance of systems.
Phoneme durations from natural singing voices were used while performing the objective evaluation.
Mean opinion score (MOS) tests were also conducted to subjectively evaluate the naturalness of synthesized waveforms.
In the subjective evaluation, the phoneme durations were determined by \textbf{DNN+ML} in Section~\ref{sec:exp-timing}.
The subjects were eleven Japanese students in our research group, and twelve phrases were chosen at random per method from the test set.
The scale for the MOS test was 5 for ``natural'' and 1 for ``poor.''
The demo songs can be found on the demo page~\cite{sinsy-taslp-demo}.

\begin{figure}[t]
  \centering
  \includegraphics[scale=0.9]{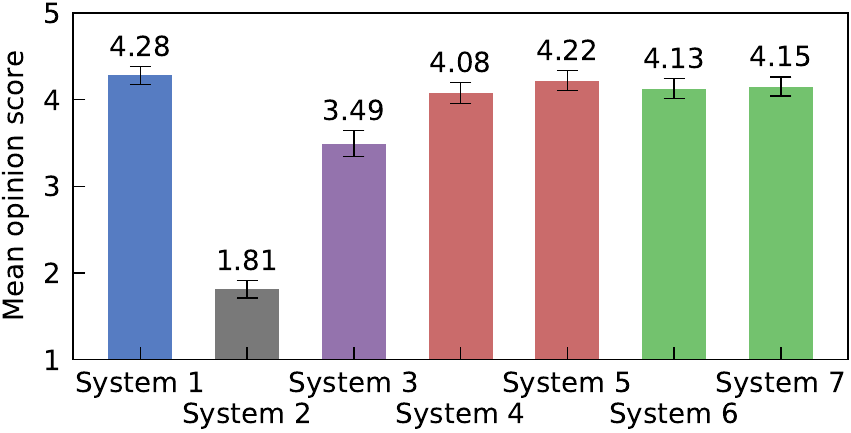}
  \vspace{-2mm}
  \caption{Mean opinion scores of the seven SVS systems with 95\% confidence intervals.}
  \label{fig:mos-acoustic}
\end{figure}

\begin{figure*}[t]
  \centering
  \subfloat[Difference-based vibrato modeling (\textbf{System~1}). \label{fig:vib-diff}]{\includegraphics[width=0.33\hsize]{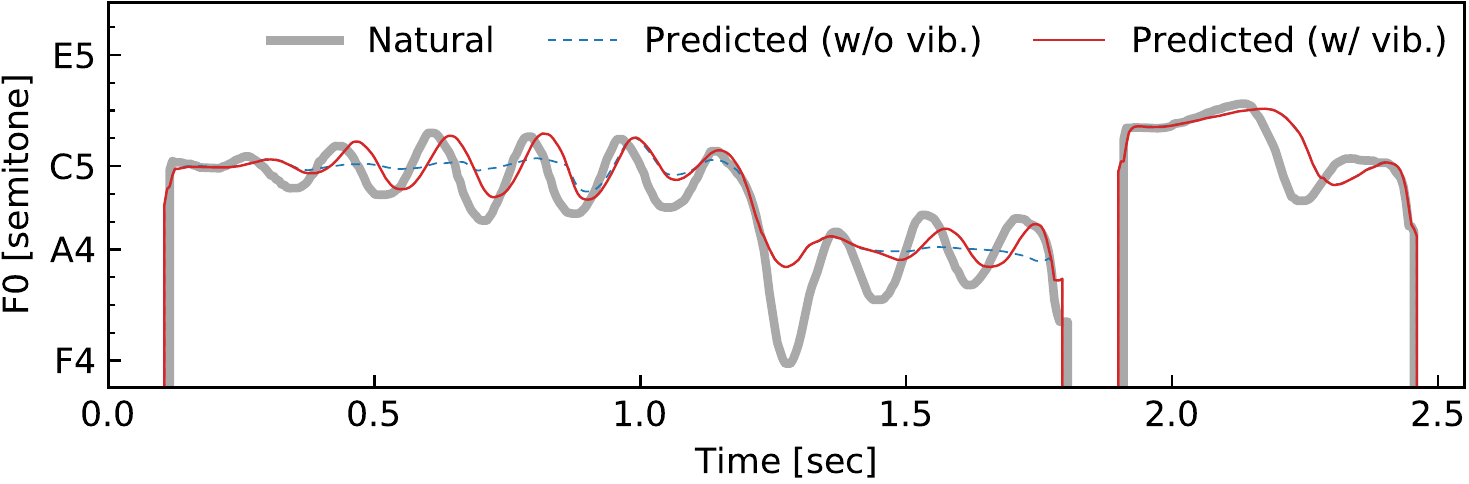}}
  \hfill
  \subfloat[Sine-based vibrato modeling (\textbf{System~4}). \label{fig:vib-sine}]{\includegraphics[width=0.33\hsize]{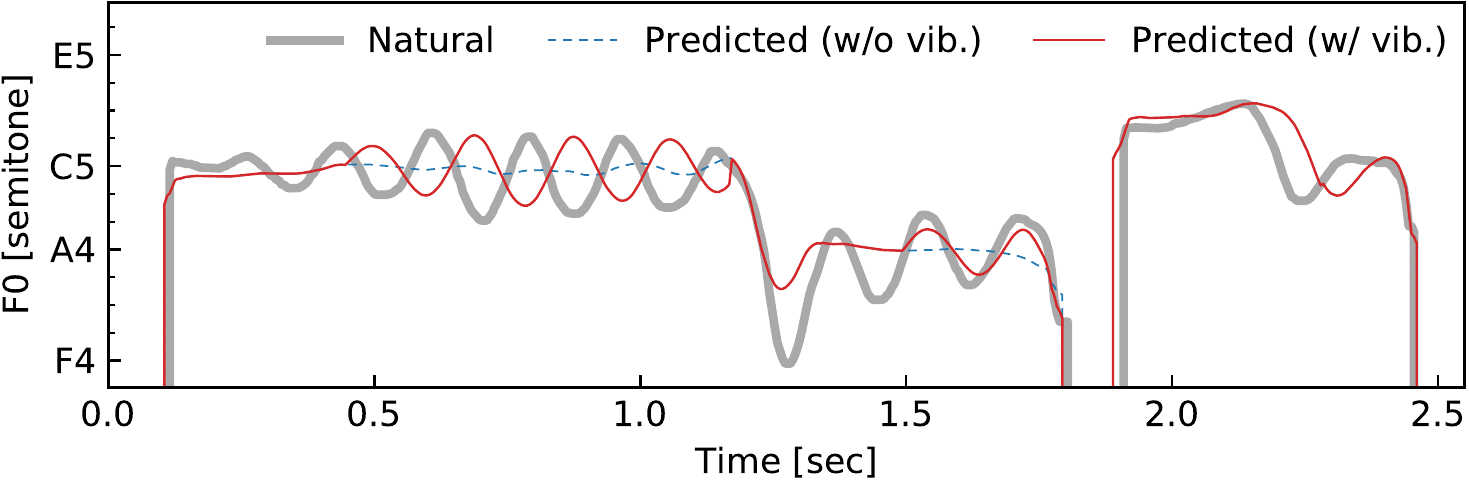}}
  \hfill
  \subfloat[Not using explicit vibrato modeling (\textbf{System~5}). \label{fig:vib-no}]{\includegraphics[width=0.33\hsize]{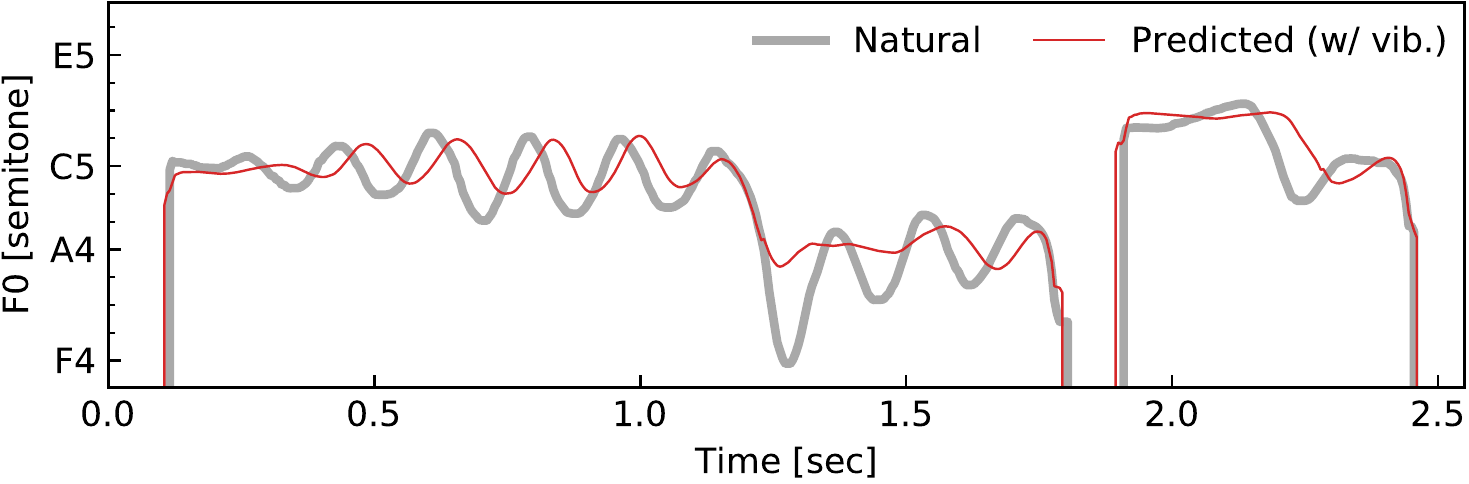}}
  \hfill
  \\
  \vspace{-3mm}
  \caption{Generated F0 contours for test song. Predicted (w/o vib.) denotes the bare F0 contour without vibrato components and Predicted (w/ vib.) denotes final F0 contour with vibrato components. The frequency value of note A4 is 440 Hz in this paper.}
  \label{fig:vib}
\end{figure*}

\begin{figure*}
  \centering
  \includegraphics[width=0.99\hsize]{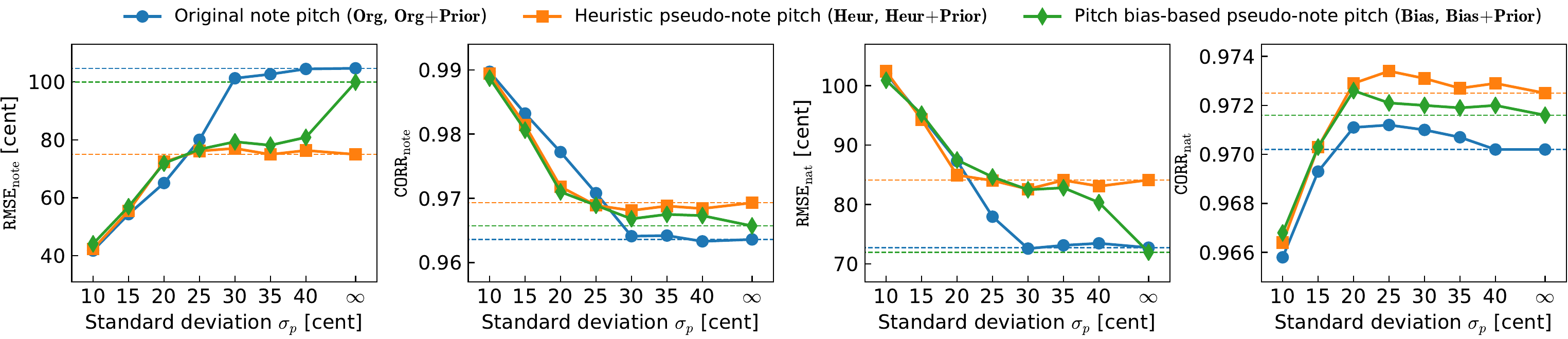}
  \vspace{-2mm}
  \caption{Objective evaluation of comparing automatic pitch correction techniques. $\prior = \infty$ denotes the systems not using prior in the training (\textbf{Org}, \textbf{Heur}, and \textbf{Bias}) and the dotted lines on each graph represent the objective evaluation values of these systems.}
  \label{fig:obj-pitch}
\end{figure*}

Table~\ref{tbl:obj-acoustic} shows the objective evaluation results and Fig.~\ref{fig:mos-acoustic} shows the subjective evaluation results.
By comparing \textbf{System~2} and \textbf{System~3} in the objective and the subjective evaluations, we revealed that the pitch normalization technique is essential in modeling the pitches of singing voices.
The synthesized singing voices in \textbf{System 2} were sometimes perceived as if they were sung following different note pitches because the generated F0 deviated from the target note pitches.
Furthermore, \textbf{System~1} achieved a better score than \textbf{System~3} in terms of both the metrics of F0 and MCD, which also led to a good subjective evaluation score.
This result suggests that it is helpful to transmit the note pitch information of the musical score to the inside of the model by using the skip connection because a singing voice is greatly affected by note pitch transition.

Comparing the methods of vibrato modeling, \textbf{System~1} and \textbf{System~5} outperformed \textbf{System~4} in terms of F0+Vib-RMSE, F0+Vib-CORR, and MOS value.
The examples of generated F0 contours in each system are plotted in Fig.~\ref{fig:vib}.
As the figure shows, the F0 contours of \textbf{System~1} and \textbf{System~5} are closer to the natural F0 contour than that of \textbf{System~4}.
Since \textbf{System~4} cannot reproduce the vibrato phase, the F0 contour deviates significantly from the natural F0 contour.
This is a major factor in the deterioration of the objective evaluation.
In addition, the start and end shapes of the vibrato of \textbf{System~1} and \textbf{System~5} are smoother than that of \textbf{System~4}, indicating the effectiveness of modeling the vibrato component by neural networks without using the sinusoidal parameters.
A comparison of \textbf{System~1} and \textbf{System~5} shows no significant difference between them.
In \textbf{System~1}, since F0 and vibrato are modeled separately, it is possible to change the vibrato intensity and introduce the pitch correction techniques described in Section~\ref{sec:pitch}.
Therefore, difference-based vibrato modeling is an effective method.

Finally, \textbf{System~1}, \textbf{System~6}, and \textbf{System~7} were compared.
\textbf{System~1} was expected to enable more continuous and appropriate parameter generation without using explicit dynamic features during synthesis because this system was trained considering the dynamic features, but the effect was slight.
Although \textbf{System~6} sometimes generated unstable singing voices, it was not a big problem in the subjective evaluation.
Our acoustic model consisted of bidirectional LSTMs that can generate sufficiently continuous parameters without using dynamic features.
Meanwhile, \textbf{System~7} may have caused parameter over-smoothing due to the parameter generation considering dynamic features explicitly.

In summary, \textbf{System~1} got a generally good objective score and achieved the best MOS value.
These results indicate the effectiveness of the proposed system with pitch normalization, the skip connection of the note pitch, difference-based vibrato modeling, and the training criterion considering dynamic features.

\subsection{Effectiveness of Automatic Pitch Correction Techniques}
\label{sec:exp-pitch}

We also evaluated the effectiveness of the automatic pitch correction techniques using a different speaker's singing voice dataset.
This dataset consisted of the same 70 songs used in the previous experiments but contained out-of-tune phrases.
Other experimental conditions were the same as in Section~\ref{sec:exp-cond}.
Time-lags and phoneme durations were modeled by \textbf{DNN+ML} in Section~\ref{sec:exp-timing}, and acoustic features were modeled by \textbf{System~1} in Section~\ref{sec:exp-acoustic}.

In this experiment, six systems were used for comparison as shown in Table~\ref{tbl:pitch-method}.
Three types of note pitches were used in the training stage: the original note pitch given by the musical scores as a baseline, a heuristic pseudo-note pitch mentioned in Section~\ref{sec:pitch-heur}, and a pitch bias-based pseudo-note pitch mentioned in Section~\ref{sec:pitch-bias}.
Note that original note pitches were always used during the synthesis stage.

\begin{table}
  \caption{Systems for Evaluation of Pitch Correction Techniques.}
  \label{tbl:pitch-method}
  \centering
  \begin{tabular}{l|cc}
    \toprule
    Note pitch & w/o prior & w/ prior \\
    \midrule\midrule
    Original note pitch & \textbf{Org} & \textbf{Org+Prior} \\
    Heuristic pseudo-note pitch & \textbf{Heur} & \textbf{Heur+Prior} \\
    Pitch bias-based pseudo-note pitch & \textbf{Bias} & \textbf{Bias+Prior} \\
    \bottomrule
  \end{tabular}
  \vspace{-2mm}
\end{table}

\begin{figure*}[t]
  \centering
  \subfloat[Heuristic pseudo-note pitch\label{fig:note-heur}]{\includegraphics[width=0.325\hsize]{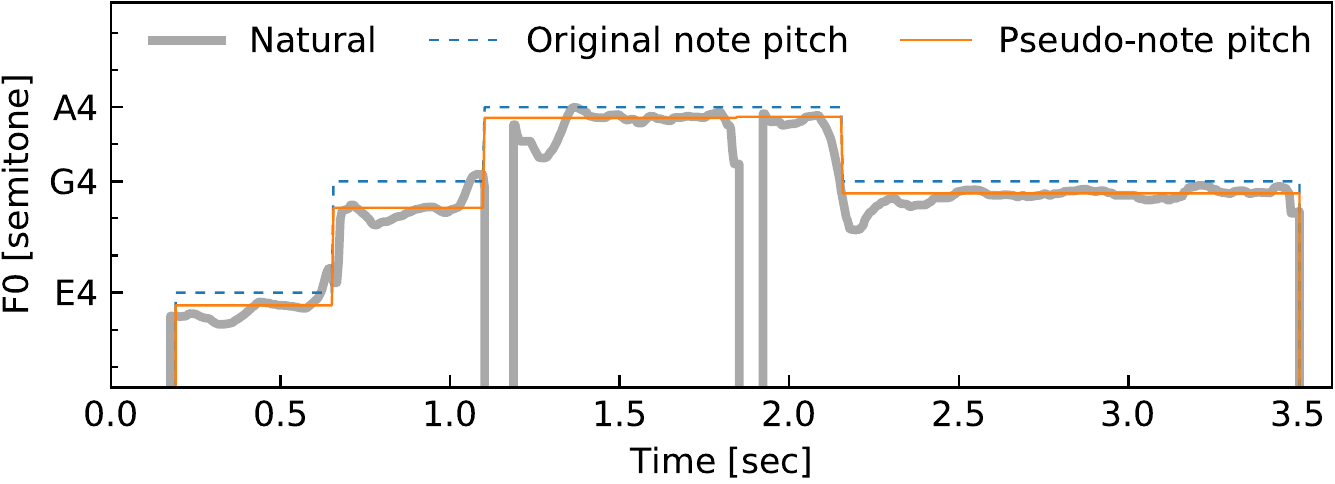}}
  \hfill
  \subfloat[Pitch bias-based pseudo-note pitch in case \textbf{Bias}. \label{fig:note-bias}]{\includegraphics[width=0.325\hsize]{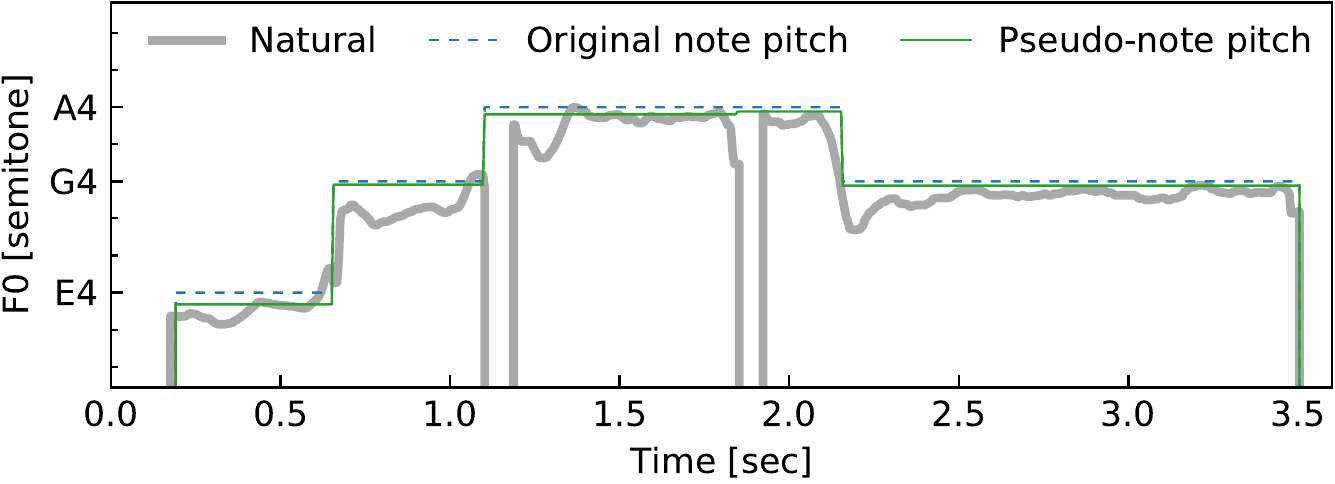}}
  \hfill
  \subfloat[Pitch bias-based pseudo-note pitch in case \textbf{Bias+Prior} ($\prior = 20$). \label{fig:note-bias-prior}]{\includegraphics[width=0.325\hsize]{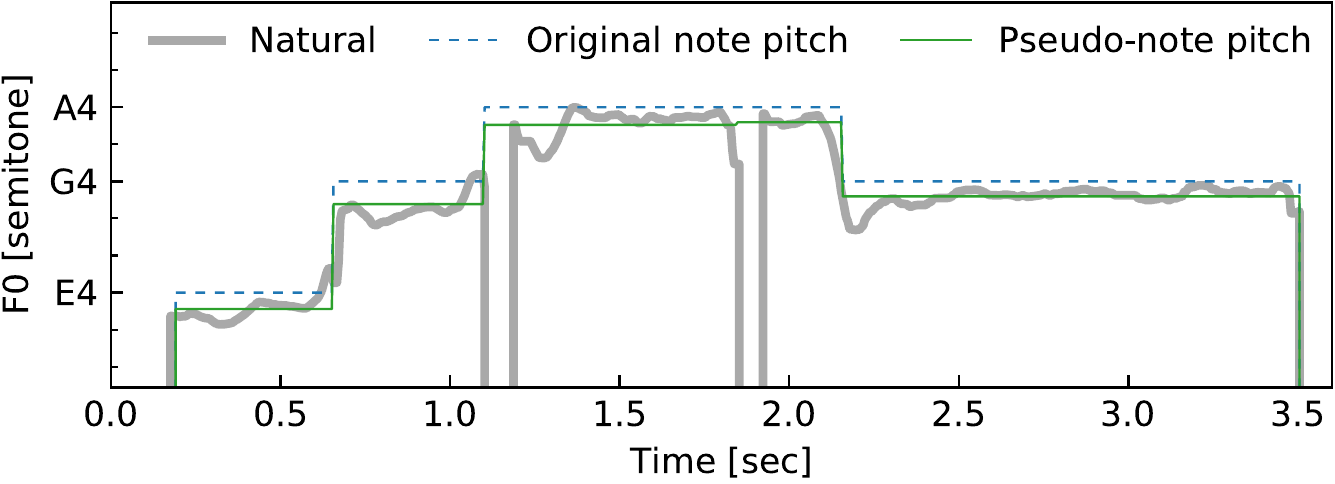}}
  \hfill
  \vspace{-1mm}
  \caption{Examples of pseudo-note pitch in proposed automatic pitch correction.}
  \label{fig:note}
  \vspace{-2mm}
\end{figure*}

\begin{figure*}[t]
  \centering
  \subfloat[\textbf{Org} \label{fig:f0-base}]{\includegraphics[width=0.325\hsize]{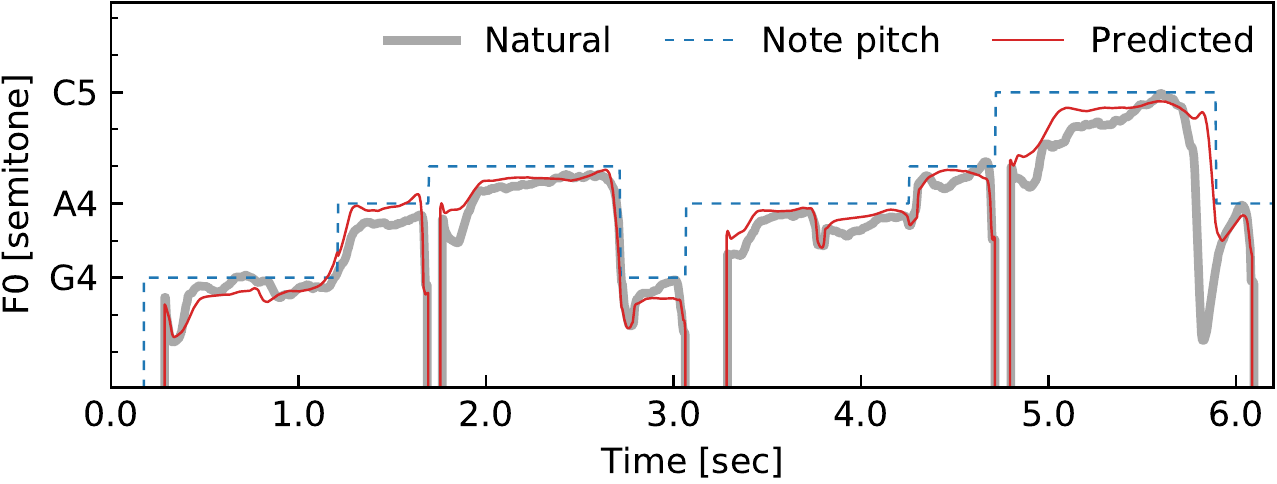}}
  \hfill
  \subfloat[\textbf{Heur} \label{fig:f0-heur}]{\includegraphics[width=0.325\hsize]{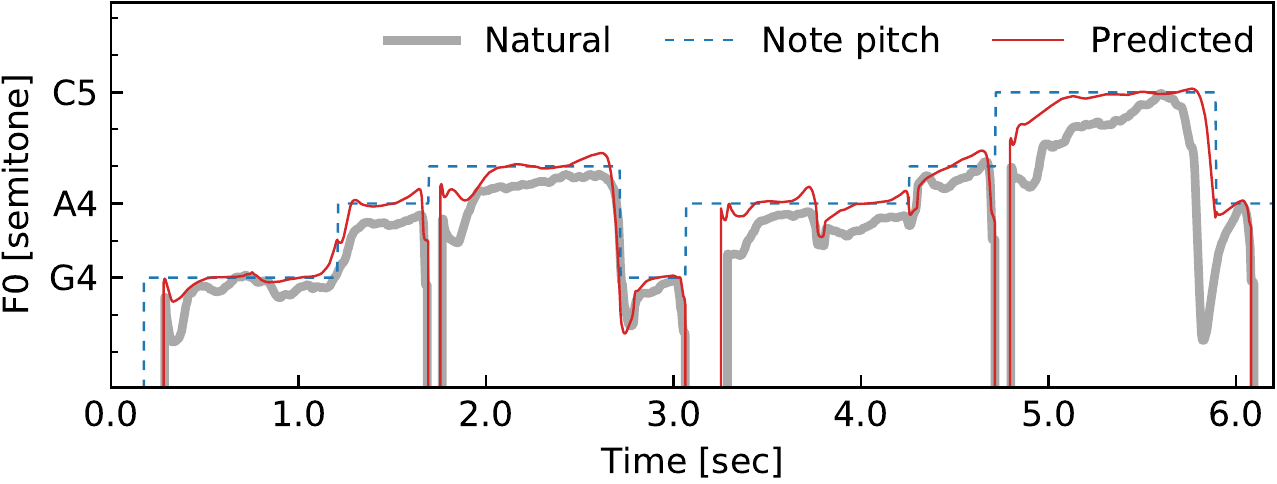}}
  \hfill
  \subfloat[\textbf{Bias} \label{fig:f0-bias}]{\includegraphics[width=0.325\hsize]{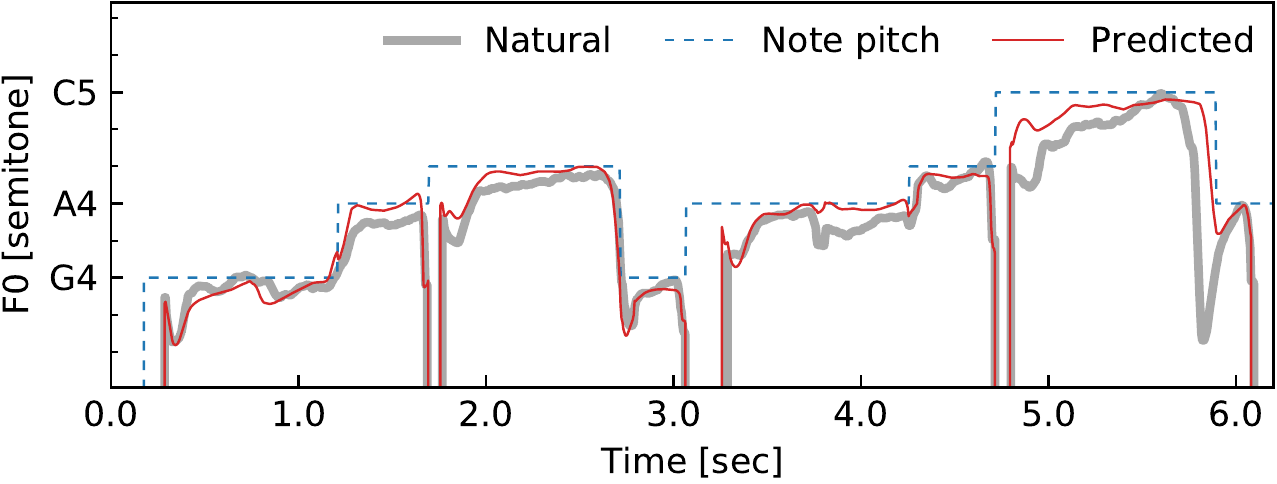}}
  \hfill
  \\
  \vspace{-2mm}
  \subfloat[\textbf{Org+Prior} ($\prior = 20$) \label{fig:f0-base-prior}]{\includegraphics[width=0.325\hsize]{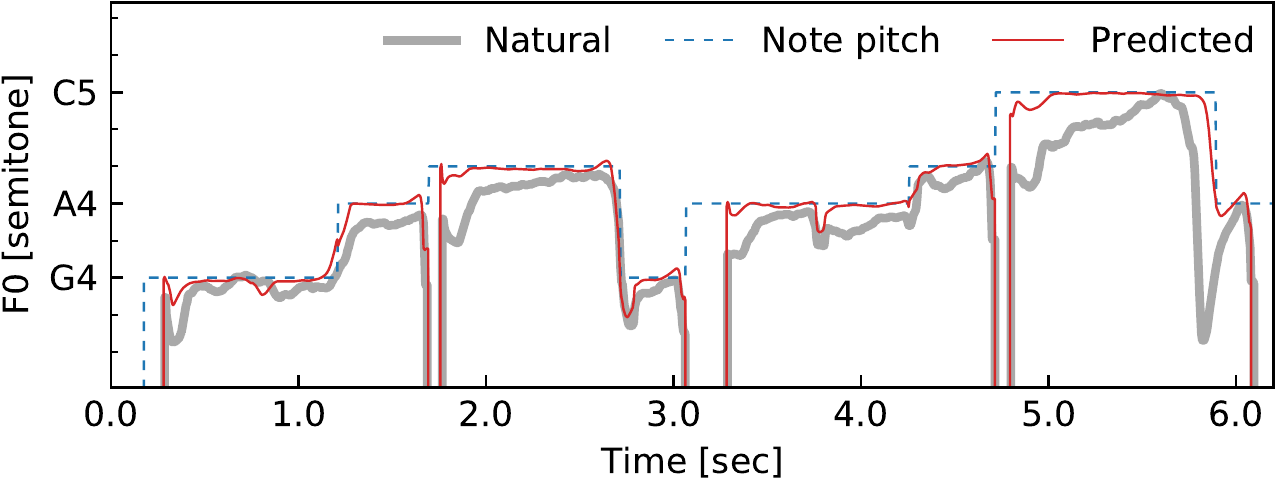}}
  \hfill
  \subfloat[\textbf{Heur+Prior} ($\prior = 20$) \label{fig:f0-heur-prior}]{\includegraphics[width=0.325\hsize]{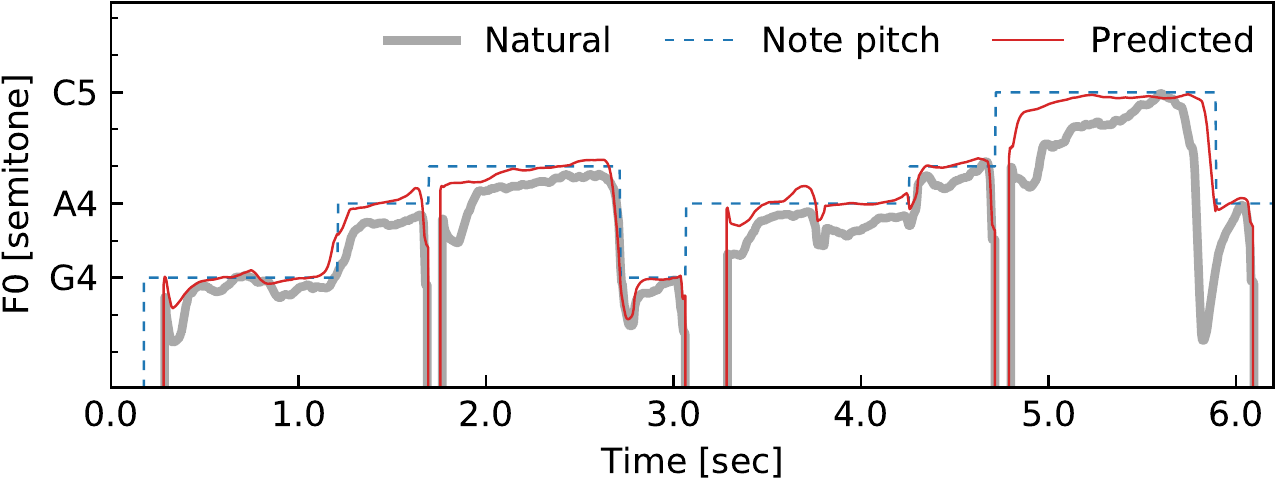}}
  \hfill
  \subfloat[\textbf{Bias+Prior} ($\prior = 20$) \label{fig:f0-bias-prior}]{\includegraphics[width=0.325\hsize]{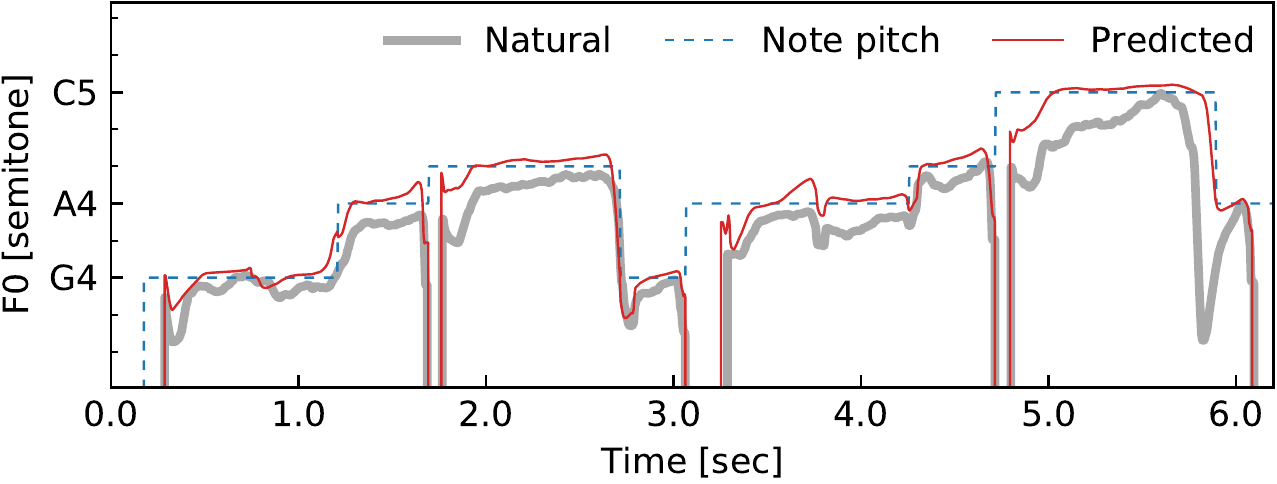}}
  \hfill
  \\
  \caption{Generated F0 contours for one test phrase.}
  \label{fig:f0}
\end{figure*}

\subsubsection{Objective Evaluation}

We compared the performance of the experimental systems objectively.
In \textbf{Org+Prior}, \textbf{Heur+Prior}, and \textbf{Bias+Prior}, we compared seven different values of standard deviation $\prior$ [cent] in \eqref{eq:loss-dpitch}.
We use four objective measures: the RMSE and CORR between the generated F0 and the F0 calculated from the note pitches (\notermse and \notecorr), and those between the generated F0 and the F0 extracted from the natural waveform (\formse and \focorr).
RMSEs and CORRs are objective measures that represent how close the value and shape of the predicted sequence are to the target sequence.
Note that if the target sequence includes out-of-tune phrases, it is not necessarily good that the \formse achieves small and the \focorr achieves high.
A small \notermse and high \notecorr mean that the generated F0 is close to the stair-like correct note pitch.

Fig.~\ref{fig:obj-pitch} shows the results of the objective evaluations.
In \textbf{Org+Prior}, by setting $\prior$ smaller than $30$, \notermse significantly decreased compared to \textbf{Org}.
This result indicates that introducing prior distributions corrected the pitch.
On the other hand, the \notecorr went higher at the same time as the improvement of \notermse, indicating that the shape of the generated F0 in \textbf{Org+Prior} tends to be a stair-like note pitch.
The results of \formse at $\prior < 30$ in \textbf{Org+Prior} also got worse because the test data for evaluation also included out-of-tune phrases.
When $\prior$ was set in the range of $20$ to $35$, \focorr in \textbf{Org+Prior} had a better score than that in \textbf{Org}.
This is because it could suppress the generation of unstable pitch fluctuations, which can be seen in the out-of-tune training data.

Compared \textbf{Heur} with \textbf{Org}, \notermse was greatly improved by introducing a heuristic pseudo-note pitch.
In addition, when combined with prior distributions in \textbf{Heur+Prior}, \notermse and \notecorr did not change if $\prior = 20$ or more.
These results show that the heuristic pseudo pitch is effective for pitch correction.
Furthermore, \textbf{Heur} and \textbf{Heur+Prior} achieve higher \focorr than \textbf{Org} and \textbf{Org+Prior}, indicating that the deviation between the original F0 and note pitch in \textbf{Heur} and \textbf{Heur+Prior} becomes smaller by using the pseudo-note pitch, thus avoiding forced pitch correction.

The results of \textbf{Bias+Prior} show a similar trend to those of \textbf{Heur+Prior}.
In contrast, the objective results of \textbf{Bias} were not as good as those of \textbf{Heur}.
The examples of heuristic pseudo-note pitch and pitch bias-based pseudo-note pitches in both \textbf{Bias} and \textbf{Bias+Prior} are shown in Fig.~\ref{fig:note}.
The pitch bias-based pseudo-note pitch of \textbf{Bias+Prior} in Fig.~\ref{fig:note}\subref{fig:note-bias-prior} is similar to the heuristic pseudo-note pitch in Fig.~\ref{fig:note}\subref{fig:note-heur}.
However, the pitch bias-based one sometimes yields inappropriate results, as shown at around 1.5 seconds in Fig.~\ref{fig:note}\subref{fig:note-bias-prior}.
This result is because the pitch bias is determined by considering the entire note and is influenced by the singing expression that changes F0 within the note, such as bending and hiccups.
On the other hand, the pitch bias-based pseudo-note pitch of \textbf{Bias} in Fig.~\ref{fig:note}\subref{fig:note-bias} is close to the original note pitch in the musical score.
This result indicates that there is ambiguity as to whether the average pitch shift of F0 at each note should be represented by the outputs of the acoustic model or the pitch biases.

\subsubsection{Subjective Evaluation}

We conducted the MOS test to evaluate the overall naturalness.
Subjects were instructed to give high score values to test phrases that were naturally pitch-corrected and not out of tune.
We compared the six systems listed in Table~\ref{tbl:pitch-method}.
The $\prior$ in \textbf{Org+Prior}, \textbf{Heur+Prior}, and \textbf{Bias+Prior} was set to $20$.
\begin{figure}
  \centering
  \includegraphics[scale=0.9]{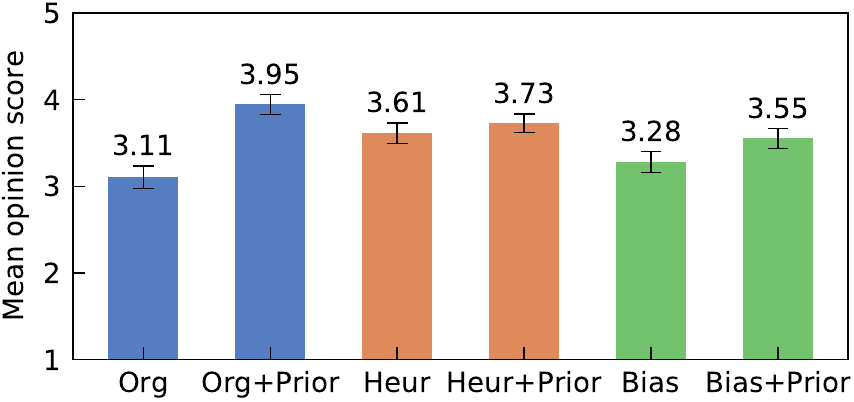}
  \vspace{-2mm}
  \caption{
  Results of MOS test comparing automatic pitch correction techniques with 95\% confidence intervals.
  In \textbf{Org+Prior}, \textbf{Heur+Prior}, and \textbf{Bias+Prior}, $\prior$ were set to $20$.
  }
  \vspace{-2mm}
  \label{fig:mos-pitch}
\end{figure}
The generated F0 contours are plotted in Fig.~\ref{fig:f0}, and the results of the MOS test are plotted in Fig.~\ref{fig:mos-pitch}.
\textbf{Org+Prior} achieved a higher MOS value than \textbf{Org}.
This result confirmed the effectiveness of using prior distribution.
\textbf{Heur} and \textbf{Heur+Prior} outperformed \textbf{Org}, but there was little difference between \textbf{Heur} and \textbf{Heur+Prior}.
Since the values of the F0 differences modeled by the acoustic model were smaller on average by introducing the heuristic pseudo pitch, the prior distribution seemed to have a limited effect.
On the other hand, although \textbf{Bias} and \textbf{Bias+Prior} also outperformed \textbf{Org}, these did not reach \textbf{Heur} and \textbf{Heur+Prior}, respectively.
As shown in Fig.~\ref{fig:f0}, \textbf{Bias} and \textbf{Bias+Prior} were more likely to generate an F0 that slightly deviates from the correct note pitch, compared with \textbf{Heur} and \textbf{Heur+Prior}.
This result indicates that it is not easy to automatically obtain pseudo-note pitches from F0, which includes fluctuations such as bending and hiccups, and the heuristic method of obtaining pseudo-note pitches is powerful and effective.
Overall, \textbf{Org+Prior} achieved the best MOS value even though \textbf{Org+Prior} showed a worse \focorr than \textbf{Heur+Prior} and \textbf{Bias+Prior}.
The objective evaluation results and Fig.~\ref{fig:f0}\subref{fig:f0-base-prior} show that the F0 contour generated by \textbf{Org+Prior} was the most similar to a stair-like F0 contour, and the output of the acoustic model in terms of the F0 was strongly corrected by the prior distribution in the training stage.
While fine fluctuations of F0 were lost, the unstable pitch fluctuation seen in the out-of-tune phrases was also suppressed, leading to good subjective evaluation results.
Appropriate correction methods and standard deviation of prior $\prior$ should be selected based on which to prioritize, the reproducibility of F0 fluctuations particular to singers in training data, or the accuracy of the pitch.
Note that none of the systems reached the subjective score of \textbf{System 1} in Section~\ref{sec:exp-acoustic}.
This result indicates that it is difficult to predict the pitch accurately even if the pitch correction technique is used when the training data contains the out-of-tune phrase.
This led to a decrease in MOS score because the naturalness of pitch fluctuation significantly affects the subjective quality of singing voices.

\section{Conclusion}
\label{sec:conclusion}

We proposed a DNN-based SVS system called ``Sinsy,'' designed to synthesize singing voice with singing-specific expressions at appropriate timing from a musical score.
The proposed system consists of four DNN-based modules: a time-lag model, a duration model, an acoustic model, and a vocoder.
The proposed system incorporates improved pitch and vibrato modeling, the better training criterion, and the pitch robust neural vocoder PeriodNet.
Furthermore, we propose pitch correction techniques that enable synthesizing singing voices with the correct pitch even if the training data has out-of-tune phrases.
Experimental results indicated the effectiveness of our novel techniques.
Our proposed system can synthesize high-quality, high-fidelity singing voices that can follow a given musical score.

Future work includes investigating different distributions for prior in automatic pitch correction and evaluating the proposed system using the different speaker, genres, and language datasets.
In our previous work~\cite{saino-2012-rap,nakamura-2014-hmm}, since an HMM-based SVS system, which had a similar strategy combining the time-lag model, duration model, and acoustic models, was applied for synthesizing singing voice with other styles and languages, we think the proposed DNN-based SVS system can also support these kinds of songs.
The modeling of songs in unique singing styles such as shouting and growling, which are difficult to annotate songs and extract acoustic feature representations, is also included in our future work.
Furthermore, incorporating our proposed techniques, such as time-lag and vibrato modeling and automatic pitch correction, into seq-to-seq SVS systems is an important task.
Recent studies~\cite{lee-2019-adversarially,angelini-2020-singing,blaauw-2020-sequence,gu-2020-bytesing,lu-2020-xiaoicesing,chen-2020-hifisinger,shi-2020-sequence} introduce a seq-to-seq model into the SVS system.
Although such systems can model the singing voice as sequential mapping using an encoder-decoder model with an attention mechanism, they cannot model and control timing fluctuation explicitly.
Extending a unified framework for simultaneously modeling acoustic feature and duration parameters~\cite{tokuda-2016-temporal} is one of our future works to model time-lags, durations, and acoustic features simultaneously.

% if have a single appendix:
%\appendix[Proof of the Zonklar Equations]
% or
%\appendix  % for no appendix heading
% do not use \section anymore after \appendix, only \section*
% is possibly needed

% use appendices with more than one appendix
% then use \section to start each appendix
% you must declare a \section before using any
% \subsection or using \label (\appendices by itself
% starts a section numbered zero.)
%

% \appendices
% \section{Proof of the First Zonklar Equation}
% Appendix one text goes here.
%
% % you can choose not to have a title for an appendix
% % if you want by leaving the argument blank
% \section{}
% Appendix two text goes here.

% use section* for acknowledgment
\section*{Acknowledgment}

% The authors would like to thank...
This work was supported by JSPS KAKENHI Grant Number JP18K11163 and CASIO SCIENCE PROMOTION FOUNDATION.

% Can use something like this to put references on a page
% by themselves when using endfloat and the captionsoff option.
\ifCLASSOPTIONcaptionsoff
  \newpage
\fi

% trigger a \newpage just before the given reference
% number - used to balance the columns on the last page
% adjust value as needed - may need to be readjusted if
% the document is modified later
%\IEEEtriggeratref{8}
% The "triggered" command can be changed if desired:
%\IEEEtriggercmd{\enlargethispage{-5in}}

% references section

% can use a bibliography generated by BibTeX as a .bbl file
% BibTeX documentation can be easily obtained at:
% http://mirror.ctan.org/biblio/bibtex/contrib/doc/
% The IEEEtran BibTeX style support page is at:
% http://www.michaelshell.org/tex/ieeetran/bibtex/
%\bibliographystyle{IEEEtran}
% argument is your BibTeX string definitions and bibliography database(s)
%\bibliography{IEEEabrv,../bib/paper}
%
% <OR> manually copy in the resultant .bbl file
% set second argument of \begin to the number of references
% (used to reserve space for the reference number labels box)
\bibliographystyle{IEEEtran}
\bibliography{references}

% \begin{thebibliography}{1}
%
% \bibitem{IEEEhowto:kopka}
% H.~Kopka and P.~W. Daly, \emph{A Guide to \LaTeX}, 3rd~ed.\hskip 1em plus
%   0.5em minus 0.4em\relax Harlow, England: Addison-Wesley, 1999.
%
% \end{thebibliography}

% biography section
%
% If you have an EPS/PDF photo (graphicx package needed) extra braces are
% needed around the contents of the optional argument to biography to prevent
% the LaTeX parser from getting confused when it sees the complicated
% \includegraphics command within an optional argument. (You could create
% your own custom macro containing the \includegraphics command to make things
% simpler here.)
%\begin{IEEEbiography}[{\includegraphics[width=1in,height=1.25in,clip,keepaspectratio]{mshell}}]{Michael Shell}
% or if you just want to reserve a space for a photo:

% \begin{IEEEbiography}{Michael Shell}
% Biography text here.
% \end{IEEEbiography}
%
% % if you will not have a photo at all:
% \begin{IEEEbiographynophoto}{John Doe}
% Biography text here.
% \end{IEEEbiographynophoto}
%
% % insert where needed to balance the two columns on the last page with
% % biographies
% %\newpage
%
% \begin{IEEEbiographynophoto}{Jane Doe}
% Biography text here.
% \end{IEEEbiographynophoto}

% You can push biographies down or up by placing
% a \vfill before or after them. The appropriate
% use of \vfill depends on what kind of text is
% on the last page and whether or not the columns
% are being equalized.

%\vfill

% Can be used to pull up biographies so that the bottom of the last one
% is flush with the other column.
%\enlargethispage{-5in}

\begin{IEEEbiography}[{\includegraphics[width=1in,height=1.25in,clip,keepaspectratio]{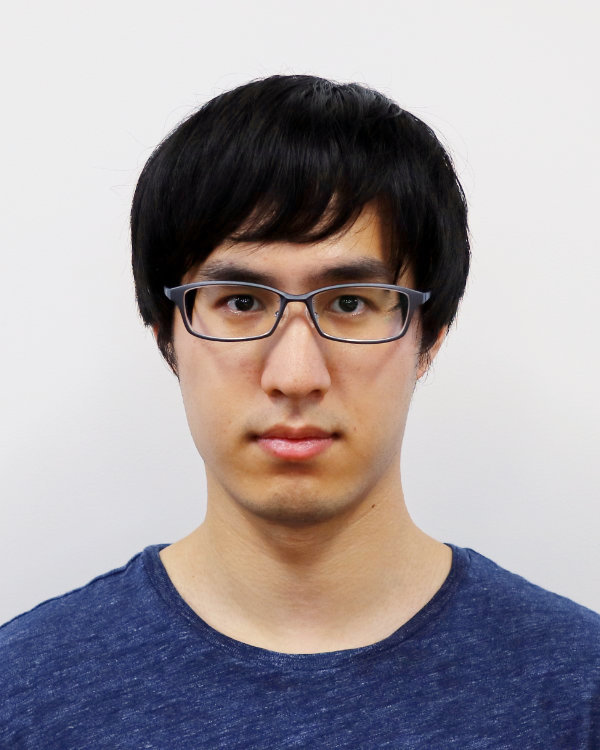}}]{Yukiya Hono}
received the B.E. and M.E. degrees in Computer Science from Nagoya Institute of Technology, Nagoya, Japan, in 2017 and 2019, respectively.
He is currently a Ph.D. candidate at Nagoya Institute of Technology.
From July to August 2019, he was an intern at Microsoft Development, Japan.
He was a visiting researcher at the University of Edinburgh, U.K., from October 2019 to December 2019 and at the University of Sheffield, U.K., from January 2020 to February 2020.
His research interests include statistical speech synthesis, singing voice synthesis, and machine learning.
He is a member of the Acoustical Society of Japan (ASJ).
He received the 18th Student Presentation Award from ASJ, the 2019 Information and Communication Engineers (IEICE) Tokai Section Student Award, and the 2021 IEEE Nagoya Section Excellent student Award.
\end{IEEEbiography}

\begin{IEEEbiography}[{\includegraphics[width=1in,height=1.25in,clip,keepaspectratio]{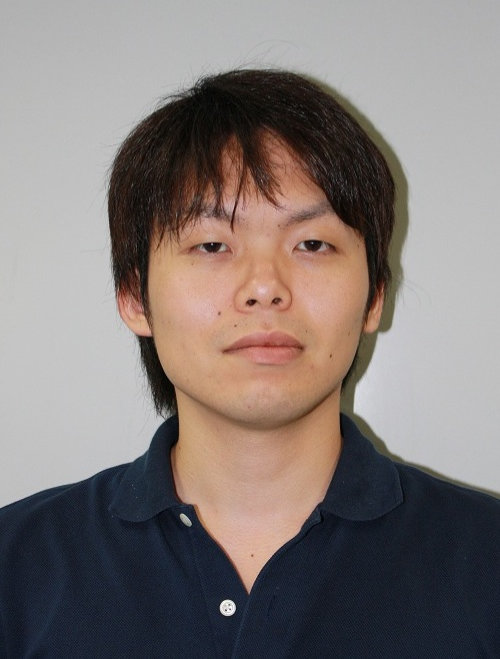}}]{Kei Hashimoto}
received the B.E., M.E., and Ph.D. degrees in Computer Science, Computer Science and Engineering, and Scientific and Engineering Simulation from Nagoya Institute of Technology, Nagoya, Japan in 2006, 2008, and 2011, respectively.
From October 2008 to January 2009, he was an intern researcher at National Institute of Information and Communications Technology (NICT), Kyoto, Japan.
From April 2010 to March 2012, he was a Research Fellow of the Japan Society for the Promotion of Science (JSPS) at Nagoya Institute of Technology, Nagoya, Japan.
From May 2010 to September 2010, he was a visiting researcher at University of Edinburgh and Cambridge University.
From April 2012 to March 2017, he was a Specially Appointed Assistant Professor at Nagoya Institute of Technology, Nagoya, Japan.
From April 2017 to December 2018, he was a Specially Appointed Associate Professor at Nagoya Institute of Technology, Nagoya, Japan, and now he is an Associate Professor at the same institute.
His research interests include statistical speech synthesis and speech recognition.
He is a member of the IEEE, the IEICE, and the Acoustical Society of Japan.
\end{IEEEbiography}

\begin{IEEEbiography}[{\includegraphics[width=1in,height=1.25in,clip,keepaspectratio]{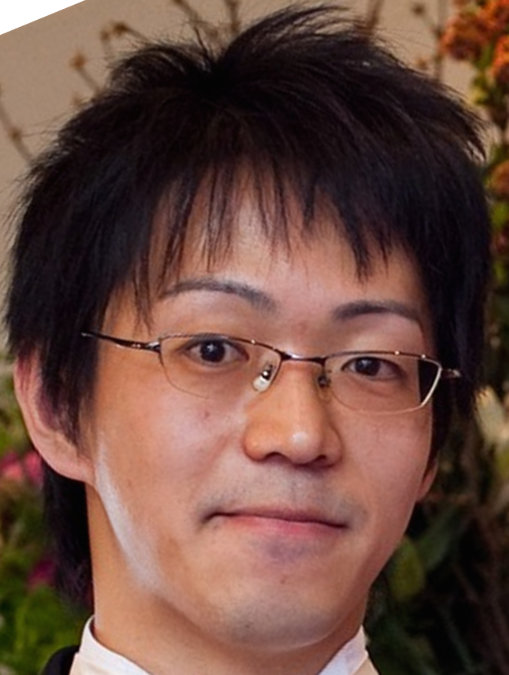}}]{Keiichiro Oura}
received his Ph.D. degree in Computer Science and Engineering from Nagoya Institute of Technology, Nagoya, Japan, in 2010.
He was a specially-appointed assistant professor at the Nagoya Institute of Technology, Japan, from Apr. 2010 to Mar. 2017.
From Apr. 2017 to May 2020, he was a specially-appointed associate professor at the Nagoya Institute of Technology.
He is currently a project associate professor at the Nagoya Institute of Technology and a CEO of the Techno-Speech, Inc.
His research interests include statistical speech recognition and synthesis.
He received the ISCSLP Best Student Paper Award, the IPSJ YAMASHITA SIG Research Award, the ASJ ITAKURA Award, the IPSJ KIYASU Special Industrial Achievement Award, the ASJ AWAYA Prize Young Researcher Award, and the IPSJ Microsoft Faculty Award, in 2008, 2010, 2013, 2013, 2019, 2020, respectively.
He is a member of the Acoustical Society of Japan and the Information Processing Society of Japan.
\end{IEEEbiography}

\begin{IEEEbiography}[{\includegraphics[width=1in,height=1.25in,clip,keepaspectratio]{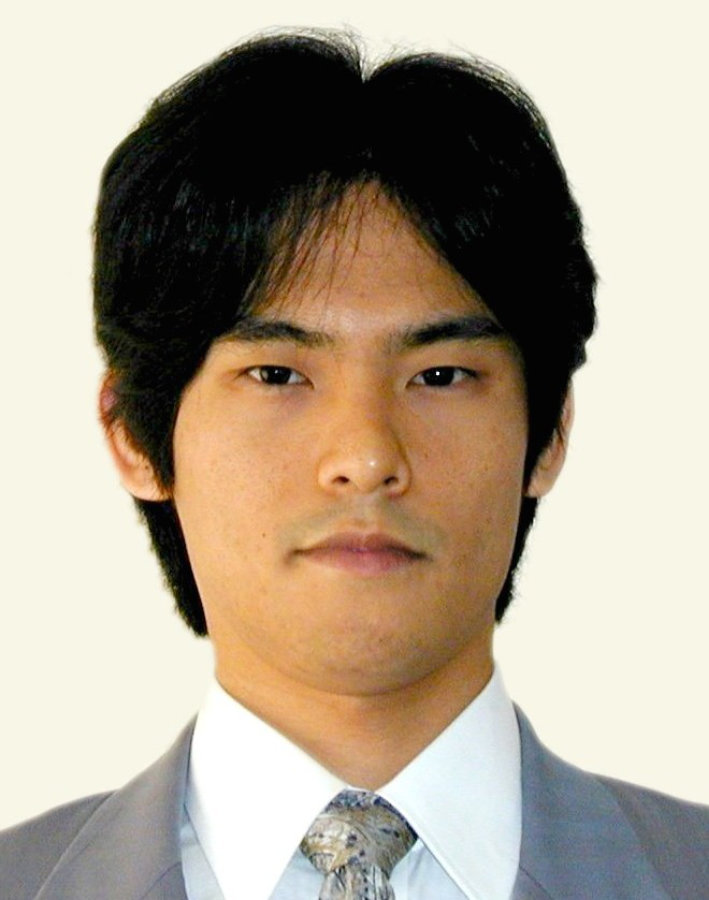}}]{Yoshihiko Nankaku}
received his B.E. degree in Computer Science, and his M.E. and Ph.D. degrees in the Department of Electrical and Electronic Engineering from Nagoya Institute of Technology, Nagoya, Japan, in 1999, 2001, and 2004 respectively.
After a year as a postdoctoral fellow at the Nagoya Institute of Technology, he became an Associate Professor at the same institute.
He was a visiting researcher at the Department of Engineering, University of Cambridge, U.K., from May to October 2011.
His research interests include statistical machine learning, speech recognition, speech synthesis, image recognition, and multi-modal interface.
He is a member of the Institute of Electronics, Information and Communication Engineers, and the Acoustical Society of Japan.
\end{IEEEbiography}

\begin{IEEEbiography}[{\includegraphics[width=1in,height=1.25in,clip,keepaspectratio]{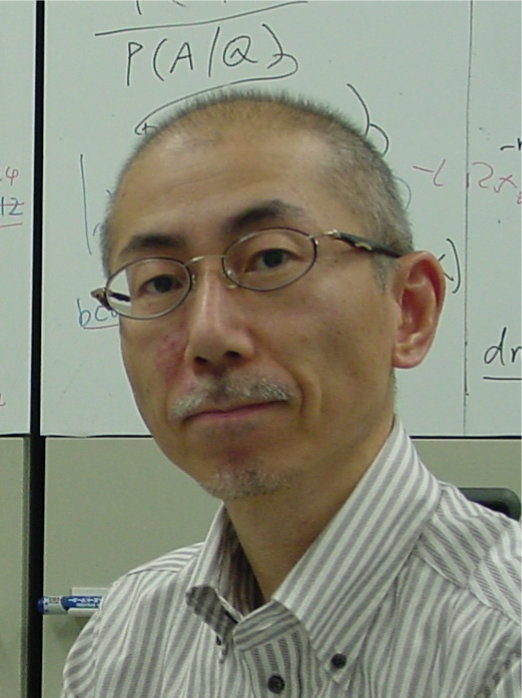}}]{Keiichi Tokuda}
received the B.E. degree in electrical and electronic engineering from Nagoya Institute of Technology, Nagoya, Japan, the M.E. and Dr.Eng. degrees in information processing from the Tokyo Institute of Technology, Tokyo, Japan, in 1984, 1986, and 1989, respectively.
From 1989 to 1996, he was a Research Associate at the Department of Electronic and Electric Engineering, Tokyo Institute of Technology.
From 1996 to 2004, he was an Associate Professor at the Department of Computer Science, Nagoya Institute of Technology as Associate Professor, and now he is a Professor at the same institute.
He is also an Honorary Professor at the University of Edinburgh.
He was an Invited Researcher at ATR Spoken Language Translation Research Laboratories, Japan, from 2000 to 2013 and was a Visiting Researcher at Carnegie Mellon University from 2001 to 2002 and at Google from 2013 to 2014.
He published over 80 journal papers and over 200 conference papers, and received six paper awards and three achievement awards.
He was a member of the Speech Technical Committee of the IEEE Signal Processing Society from 2000 to 2003, a member of ISCA Advisory Council and an associate editor of IEEE Transactions on Audio, Speech and Language Processing, and acts as organizer and reviewer for many major speech conferences, workshops, and journals.
He is an IEEE Fellow and ISCA Fellow.
His research interests include speech coding, speech synthesis and recognition, and statistical machine learning.
\end{IEEEbiography}

% that's all folks
\end{document}